\documentclass[11pt]{article}
\usepackage{graphicx}
\usepackage[margin=1in]{geometry}
\usepackage{caption}
\usepackage{subcaption}
\usepackage{amsmath}
\usepackage{amssymb}
\usepackage{color}

\begin{document}

\begin{titlepage}
\begin{center}

{ \huge \bfseries Signatures of non-gaussianity in the isocurvature modes of primordial black hole dark matter}\\[1cm]

Sam Young$^{1}$, Christian T. Byrnes$^{2}$\\[0.5cm]
Department of Physics and Astronomy, Pevensey II Building, University of Sussex, BN1 9RH, UK\\[0.5cm]
$^{1}$S.M.Young@sussex.ac.uk, $^{2}$C.Byrnes@sussex.ac.uk \\[1cm]

\today\\[1cm]

\end{center}

Primordial black holes (PBHs) are black holes which may have formed very early on during the radiation dominated era in the early universe. 
We present here a method by which the large scale perturbations in the density of primordial black holes may be used to place tight constraints on non-gaussianity if PBHs account for dark matter (DM). The presence of local-type non-gaussianity is known to have a significant effect on the abundance of primordial black holes, and modal coupling from the observed CMB scale modes can significantly alter the number density of PBHs that form within different regions of the universe, which appear as DM isocurvature modes. Using the recent \emph{Planck} constraints on isocurvature perturbations, we show that PBHs are excluded as DM candidates for even very small local-type non-gaussianity, $|f_{NL}|\approx0.001$ and remarkably the constraint on $g_{NL}$ is almost as strong. Even small non-gaussianity is excluded if DM is composed of PBHs. If local non-Gaussianity is ever detected on CMB scales, the constraints on the fraction of the universe collapsing into PBHs (which are massive enough to have not yet evaporated) will become much tighter.

\end{titlepage}

\tableofcontents

\section{Introduction}
Primordial black holes (PBHs) are black holes which theoretical arguments suggest might have formed from the direct gravitational collapse of large density perturbations very shortly after the end of inflation. PBHs may theoretically form with any mass, although their abundance is typically well constrained by observations. Whilst PBHs with mass lower than $10^{15}$g would have evaporated by today (with the possible exception of Planck mass relics), more massive PBHs would still survive, and represent a viable dark matter (DM) candidate.

Many efforts have been made to observe PBHs, and whilst they have not yet been seen, this has led to many corresponding constraints on their abundance in different mass ranges \cite{Carr:2009jm}. The econstraints typically assume that PBHs form at a single mass scale and are stated in terms of the mass fraction of the universe going into PBHs at the time of formation, $\beta$. There exists only a narrow window in which PBHs of a single mass could make up the entirety of DM, with other scales being excluded by observations. It is noted that there has been a recent claim that the tidal capture of PBHs by neutron stars could be used to exclude the remaining window (apart from Planck mass remnants) \cite{Pani:2014rca}, but this has been refuted in \cite{Capela:2014qea,Defillon:2014wla}. The results presented here can also be applied if DM is composed of smaller PBHs which have all but evaporated by today leaving Planck mass remnants which may make up DM \cite{Carr:1994ar}. Whilst this mass range is not explicitly considered, it is certainly not ruled out by observations, and the results presented here are almost independent of the PBH mass.

In order for a significant number of PBHs to form, the power spectrum on small scales needs to be significantly larger than observed in the CMB - of order $10^{-2}$ in the case of gaussian perturbations. This is possible in many models of inflation, including the running mass model \cite{Drees:2011hb}, axion inflation \cite{Bugaev:2013fya}, a waterfall transition during hybrid inflation \cite{Bugaev:2011wy, Lyth:2012yp,Halpern:2014mca}, from passive density fluctuations \cite{Lin:2012gs}, or in inflationary models with small field excursions but which are tuned to produce a large tensor-to-scalar ratio on large scales \cite{Hotchkiss:2011gz}. See also \cite{Linde:2012bt,Torres-Lomas:2014bua,Suyama:2014vga}, and a summary of various models which can produce PBHs is presented in \cite{Green:2014faa}. Alternatively, the constraint on the formation criteria can be relaxed during a phase transition in the early universe, causing PBHs to form preferentially at that mass scale \cite{Jedamzik:1999am} - although such an effect will not be considered here.

PBHs have traditionally been used to investigate the early universe by placing a constraint on the small scale power spectrum from the corresponding constraint on their abundance \cite{Green:1997sz,Josan:2009qn,Shandera:2012ke}. In this paper, large scale fluctuations in the PBH density caused by local-type non-gaussianity are considered. If DM is composed entirely, or partially, of PBHs, these perturbations will be seen as isocurvature modes in cold dark matter (CDM) - upon which there are tight constraints from the recent \emph{Planck} data release \cite{Ade:2015lrj}.

The isocurvature perturbations are formed in a highly non-linear manner in this model. PBHs form shortly after horizon reentry during radiation domination, with an energy density exponentially sensitive to the amplitude of the power spectrum. Observational constraints imply that at most one region in a million collapsed into a PBH so the large scale radiation density is almost unaffected, but if PBHs form DM then the amplitude of the DM perturbation is extremely sensitive to the modal coupling. Using this mechanism, CDM (with zero pressure) is formed in a universe which could have previously have been made up entirely of radiation and hence had no isocurvature perturbation prior to PBH formation. Such an effect is impossible within linear perturbation theory \cite{Wands:2002bn}.

In a previous paper,  the peak-background split was used to investigate the effect of modal coupling on the constraints which can be placed on the small scale power spectrum \cite{Young:2014oea}. In this paper we use the same mechanism to investigate the extent to which modal coupling produces CDM isocurvature modes and discuss the implications of such an effect. Even if the initial conditions are adiabatic, which has been shown to be the case in single-field inflation, if there is modal coupling then the conversion of radiation into CDM (by collapse into PBHs) can have different efficiencies in different regions of the universe, which introduces isocurvature modes in the CMD after inflation has ended.

Even single-field inflation generates a small value of $f_{NL}$ with magnitude comparable to the spectral index \cite{Maldacena:2002vr} - which apparently could therefore rule out single-field inflation as a mechanism to create PBH DM. However, it has been argued that this is a result of gauge choice \cite{Pajer:2013ana,Tanaka:2011aj}, and that for our purposes the effective $f_{NL}=0$ in single-field inflation. It is therefore assumed in this paper that $f_{NL}$ can be arbitrarily close to zero.

Throughout, we will assume $f_{NL}$ to be scale invariant whilst the power spectrum becomes several orders of magnitude larger at small scales - which is likely to be unrealistic given a specific model. However, this is a conservative approach, because if $|f_{NL}|$ were to become larger at some small scale, it would not weaken the constraints derived here, but would be likely to strengthen them. Even if the bispectrum was exactly zero when all three modes have sub CMB scales, the modal coupling between the CMB and PBH scales would still effect the amplitude of the power spectrum on PBH scales and the constraints which we derive would not be significantly weakened. In such a case, the perturbations within a region smaller than we can probe on the CMB would be Gaussian, but the variance would vary between different patches, in a way completely correlated to the long wavelength perturbation.   

Shortly prior to the release of this paper, Tada and Yokoyama \cite{Tada:2015noa} released a paper discussing a similar effect and the use of PBHs as biased tracers. We confirm their results and extend the calculation to account for the non-gaussianity parameter $g_{NL}$ as well as $f_{NL}$, the effect of intermediate modes (between the CMB- and PBH-scales), and make use of the more recent results from the \emph{Planck} 2015 data release. Because all surviving PBHs necessarily behave as at least a subdominant DM component today, we also show how the allowed fraction of PBHs can be constrained more tightly than previously realised, under the presence of even small non-Gaussianity.

The layout of this paper is as follows: in section 2, the calculation of the PBH abundance, in both the gaussian and non-gaussian case, is reviewed. In section 3, modal coupling and how the peak-background split may be used to investigate its effects on PBH abundance is discussed. In section 4, the calculation is applied to the formation of CDM isocurvature modes and place constraints on the non-gaussianity parameters in the case of PBH DM, and the calculation is extended to include simultaneous $f_{NL}$ and $g_{NL}$, intermediate modes, and the case where PBHs only make up a portion of the DM. We conclude with a summary of our arguments in section 6.

\section{Calculating the abundance of PBHs}

The abundance of PBHs is normally stated in terms of $\beta$: the energy fraction of the universe going into PBHs at the time of formation. The standard calculation used in the literature uses a Press-Schechter approach, although it has been shown that, for a gaussian distribution, this matches well when the theory of peaks is used. It has been argued that the density contrast, rather than the curvature perturbation, should be used - although an approximation using the curvature perturbation works very well if care is taken to exclude super-horizon modes from the calculation, and this simplifies the calculation greatly. In this section, we will briefly review the calculation, as well as the main sources of error, for both gaussian and non-gaussian cases.

When a perturbation reenters the horizon, if its amplitude exceeds a certain threshold, or critical, value, then gravitational forces will overcome pressure forces and the region will collapse to form a primordial black hole. There has been extensive research to calculate the threshold value \cite{Niemeyer:1999ak,Hawke:2002rf,Musco:2004ak,Musco:2008hv,Harada:2013epa,Nakama:2013ica}, which is typically stated in terms of the density contrast. The critical value of the density perturbation is believed to be $\delta_{c}\approx0.45$. However, in this paper the curvature perturbation is used, and the corresponding critical value is $\zeta_{c}\approx1$ - within the range found by \cite{Shibata:1999zs}, and is consistent with using the density contrast \cite{Young:2014ana}.

The main source of uncertainty in the critical value is due to the unknown shape of primordial perturbations - and this is the largest source of error in the calculation of the abundance. However, whilst the effect on the calculated value of the abundance is large, the effect of this uncertainty on derived parameters is relatively small. For example, an error of $\mathcal{O}10\%$ in the threshold value results in an error of several orders of magnitude in the calculated $\beta$ but only an error of $\mathcal{O}10\%$ in the constraint on the power spectrum \cite{Byrnes:2012yx,Young:2013oia}. In this paper, because our results depend only on the relative abundance of PBHs in different regions of the universe, the conclusions are not sensitive to small changes in the threshold value.

Using a Press-Schechter approach, the mass fraction of the universe going into PBHs at the time of formation is given by integrating over the probability density function (PDF),
\begin{equation}
\beta=\int\limits_{\zeta_c}^{\infty}P(\zeta)d\zeta.
\end{equation}
In the case of a gaussian distribution, the probability density function is
\begin{equation}
P(\zeta)=\frac{1}{\sqrt{2 \pi \sigma^{2}}}\exp\left(-\frac{\zeta}{2\sigma^{2}}\right),
\end{equation}
Where $\sigma^{2}$ is the variance of perturbation amplitude at the PBH forming scale. $\beta$ can therefore be written in terms of the complimentary error,
\begin{equation}
\beta=\mathrm{erfc}\left(\frac{\zeta_{c}}{\sqrt{2\sigma^{2}}}\right).
\end{equation}
Expanding using the large-x limit of $\mathrm{erfc}(x)$, gives
\begin{equation}
\beta\approx\sqrt{\frac{2\sigma^{2}}{\pi\zeta_{c}^{2}}}\exp\left(-\frac{\zeta_{c}^{2}}{2\sigma^{2}}\right).
\end{equation}
This is valid only if the distribution is gaussian, and because PBHs form in the extreme positive tail of the PDF, their abundance is very sensitive to any non-gaussianity, which we discuss below.

\subsection{Calculating the abundance of PBHs in the presence of non-gaussianity}
In the local model of non-gaussianity, the curvature perturbation is given by
\begin{equation}
\zeta=\zeta_{G}+\frac{3}{5}f_{NL}\left(\zeta_{G}^{2}-\sigma^{2}\right)+\frac{9}{25}g_{NL}\zeta_{G}^{3}+...=h\left(\zeta_{G}\right),
\label{local NG}
\end{equation}
where $\sigma^{2}$ is the variance of the gaussian variable $\zeta_{G}$, and is subtracted to ensure the expectation value of $\zeta$ is zero.

The calculation of the abundance of PBHs is most easily performed by calculating the values of $\zeta_{G}$ which correspond the critical value, $\zeta_{c}$, and integrating over the corresponding regions of the gaussian PDF of $\zeta_{G}$ - the reader is directed to \cite{Byrnes:2012yx,Young:2013oia} for a full derivation. For example, let us consider the case where $g_{NL}$ and higher order terms are zero:
\begin{equation}
\zeta=\zeta_{G}+\frac{3}{5}f_{NL}\left(\zeta_{G}^{2}-\sigma^{2}\right)=h\left(\zeta_{G}\right).
\end{equation}
$h^{-1}(\zeta_{c})$ therefore has two solutions, given by
\begin{equation}
h_{c\pm}^{-1}=h_{\pm}^{-1}(\zeta_{c})=\frac{-5\pm\sqrt{25+60\zeta_{c}f_{NL}+36\zeta_{c}^{2}f_{NL}^{2}\sigma^{2}}}{6f_{NL}}.
\label{hfnl}
\end{equation}
For positive $f_{NL}$
\begin{equation}
\label{pos fnl beta}
\beta=\sqrt{\frac{2}{\pi\sigma^{2}}}\left(\int\limits_{h_{c+}^{-1}}^{\infty}\exp\left(-\frac{\zeta_{G}^{2}}{2\sigma^{2}}\right)d\zeta_{G}+\int\limits_{-\infty}^{h_{c-}^{-1}}\exp\left(-\frac{\zeta_{G}^{2}}{2\sigma^{2}}\right)d\zeta_{G} \right),
\end{equation}
and for negative $f_{NL}$
\begin{equation}
\label{neg fnl beta}
\begin{split}
\beta & =\sqrt{\frac{2}{\pi\sigma^{2}}}\int\limits_{h_{c+}^{-1}}^{h_{c-}^{-1}}\exp\left(-\frac{\zeta_{G}^{2}}{2\sigma^{2}}\right)d\zeta_{G}\\
& = \sqrt{\frac{2}{\pi\sigma^{2}}}\left(\int\limits_{h_{c+}^{-1}}^{\infty}\exp\left(-\frac{\zeta_{G}^{2}}{2\sigma^{2}}\right)d\zeta_{G} - \int\limits_{h_{c-}^{-1}}^{\infty}\exp\left(-\frac{\zeta_{G}^{2}}{2\sigma^{2}}\right)d\zeta_{G} \right).
\end{split}
\end{equation}
Furthermore, if we make the assumption that $f_{NL}$ is small, $f_{NL}\ll1$, which we will show is justified in the case that DM is composed of PBHs (and is further verified by the findings of \cite{Tada:2015noa}), the above expressions can  be simplified further. In the expression of $\beta$ for positive and negative $f_{NL}$, the first term inside the brackets dominates, and $\beta$ can be written in terms of one complimentary error function,
\begin{equation}
\label{beta approx}
\begin{split}
\beta & =\sqrt{\frac{2}{\pi\sigma^{2}}}\int\limits_{h_{c+}^{-1}}^{\infty}\exp\left(-\frac{\zeta_{G}^{2}}{2\sigma^{2}}\right)d\zeta_{G}\\
& =\mathrm{erfc}\left(\frac{h_{c+}^{-1}}{\sqrt{2}\sigma}\right)\\
& \approx\sqrt{\frac{2\sigma^{2}}{\pi(h^{-1}_{c+})^{2}}}\exp\left(-\frac{(h^{-1}_{c+})^{2}}{2\sigma^{2}}\right).
\end{split}
\end{equation}
Deriving an analytic expression as shown here is not a necessary step, but it is a useful approximation, and we will later use this result to derive an analytic expression for bias factor and amplitude of isocurvature modes in the PBH density.

Although it is not shown here, the same calculation can be performed for the local model of non-gaussianity containing $g_{NL}$ - the interested reader is again directed to \cite{Byrnes:2012yx,Young:2013oia} for a full discussion of the calculation. In the case where only a cubic and linear term are considered
\begin{equation}
\zeta=\zeta_{G}+\frac{9}{25}g_{NL}\zeta_{G}^{3}=h(\zeta_{G}),
\end{equation}
then $h^{-1}(\zeta_c)$ has up to three possible solutions, depending on the value of $g_{NL}$ and $\zeta_{c}$. However, assuming that $g_{NL}$ is small, $g_{NL}\ll1$, which again will be shown later, the expression is dominated by one $\mathrm{erfc}$ function as in equation (\ref{beta approx}), with a different expression for $h^{-1}(\zeta_c)$. To first order in $g_{NL}$
\begin{equation}
h^{-1}_{c}=\zeta_{c}-\frac{9\zeta_{c}^{3}g_{NL}}{25}.
\end{equation}


\section{Modal coupling and the peak-background split}
\begin{figure}[t]
\centering
\begin{subfigure}{0.8\textwidth}
	\centering
	\includegraphics[width=\linewidth]{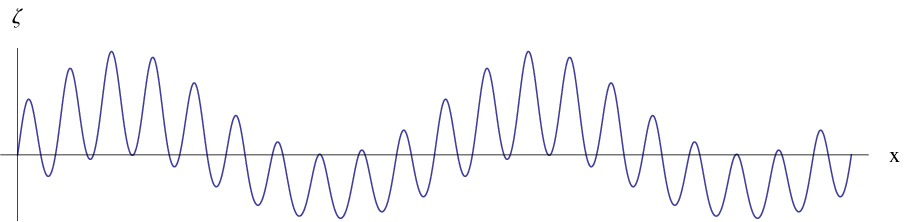}
\end{subfigure}
\begin{subfigure}{0.8\textwidth}
	\centering
	\includegraphics[width=\linewidth]{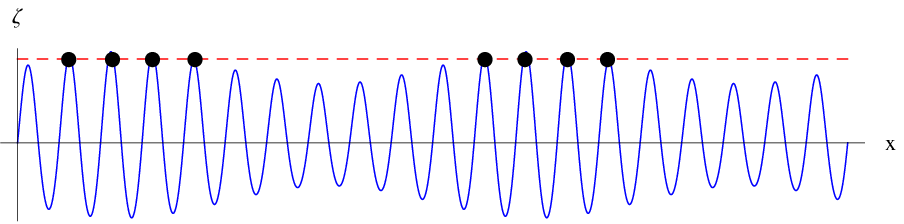}
\end{subfigure}
 \caption{The top plot shows an example of a universe containing only two modes. As an example of modal coupling, the amplitude of the short wavelength mode is a function of the long wavelength mode - the amplitude of the short-wavelngth mode is larger at the peak of the long-wavelength mode. At the time when short-wavelength mode enters the horizon, and PBHs at that scale form, the long-wavelength mode is not yet visible and will not affect whether a PBH forms or not. The bottom plot shows the same universe, but with the long wavelength mode subtracted, enabling $\zeta$ to be used a formation criterion for PBHs. The dashed red line shows the formation threshold for PBHs - regions where the curvature perturbation is greater than the formation threshold will collapse to form a PBH. The black circles represent areas which will collapse to form a PBH. It can be seen that a relatively small change in the amplitude of the small scale mode can have a large impact on the number of PBHs forming in a region.}
\label{large scale inhomos}
\end{figure}

It has previously been shown \cite{Young:2014ana} that curvature perturbation modes which are a long way outside the horizon at the time of PBH formation have little effect on whether a PBH forms. This is due to the suppression of large scale density modes by a factor $k^{2}$ relative to the curvature perturbation. In radiation domination:
\begin{equation}
\delta(t,k)=\frac{2(1+\omega)}{5+3\omega}\left(\frac{k}{aH}\right)^{2}\zeta(k)=\frac{4}{9}\left(\frac{k}{aH}\right)^{2}\zeta(k),
\end{equation}
where $\omega=1/3$ is the equation of state, and $(aH)^{-1}$ is the horizon scale at the time of PBH formation. However, long wavelength modes can have an indirect effect on the abundance of PBHs, $\beta$, due to modal coupling from non-gaussianity. A long wavelength mode can affect both the amplitude and distribution of the small scale perturbations which may form PBHs. In figure \ref{large scale inhomos}, we show how the coupling of long- and short-wavelength modes can affect the number of PBHs forming in different regions of the universe. At the peak of the long wavelength mode, the amplitude of the small scale mode is increased, forming more PBHs, whilst the opposite occurs at the trough. 

How modal coupling can affect the constraints on the power spectrum at small scales from PBHs has been investigated \cite{Young:2014oea}, although it was assumed that all the modes involved were sub-CMB and potentially had a large amplitude. In this paper, we will go beyond previous work and study the case where the large-scale modes are observable in the CMB and hence very small. Despite the their small amplitude, we show that these perturbations have a remarkably large effect on observations. In this section, we will briefly review the calculation using the peak-background split to investigate modal coupling due to the local non-gaussianity parameters $f_{NL}$ and $g_{NL}$, and in the following section, apply this to the abundance of PBHs and the creation of isocurvature modes.

\subsection{Quadratic non-gaussianity, $f_{NL}$}
We will take the model of local non-gaussianity, in terms of the curvature perturbation $\zeta$, to be described by
\begin{equation}
\label{quadratic local model}
\zeta=\zeta_{G}+\frac{3}{5}f_{NL}\left(\zeta_{G}^{2}-\sigma^{2}\right)=h(\zeta_{G}),
\end{equation}
where $\zeta_{G}$ is a gaussian variable. It is necessary to subtract $\sigma^{2}=\langle\zeta_{G}^{2}\rangle$ so that the background (average) value of $\zeta$ remains zero. We will now employ the peak-background split, and write the gaussian component as the sum of a long-(background) and short-(peak) wavelength component,
\begin{equation}
\zeta_{G}=\zeta_{l}+\zeta_{s}.
\end{equation}
Equation (\ref{quadratic local model}) then becomes:
\begin{equation}
\zeta=\left(\zeta_{l}+\zeta_{s}\right)+\frac{3}{5}f_{NL}\left(\left(\zeta_{l}+\zeta_{s}\right)^{2}-\langle\left(\zeta_{l}+\zeta_{s}\right)^{2}\rangle\right).
\end{equation}
However, terms which depend only on the long-wavelength mode do not affect PBH formation, and should not be considered when determining the abundance of PBHs. We therefore subtract those terms, leaving:
\begin{equation}
\label{quadratic inhomogeneity}
\zeta=\left(1+\frac{6}{5}f_{NL}\zeta_{l}\right)\zeta_{s}+\frac{3}{5}\left(\zeta_{s}^{2}-\sigma_{s}^{2}\right).
\end{equation}
We can now rewrite the expression in terms of new variables, $\tilde{\zeta}_{G}$, $\tilde{\sigma}$ and $\tilde{f}_{NL}$, and calculate the abundance of PBHs $\beta$ as described in section 2, as a function of the long wavelength mode, $\zeta_{l}$.
\begin{equation}
\label{local fnl variables}
\begin{split}
&\tilde{\zeta}_{G}=\left(1+\frac{6}{5}f_{NL}\zeta_{l}\right)\zeta_{s},\\
&\tilde{\sigma}=\left(1+\frac{6}{5}f_{NL}\zeta_{l}\right)\sigma_{s},\\
&\tilde{f}_{NL}=\left(1+\frac{6}{5}f_{NL}\zeta_{l}\right)^{-2}f_{NL}.\\
\end{split}
\end{equation}
Equation (\ref{quadratic inhomogeneity}) can then be written in a form analogous to equation (\ref{quadratic local model}),
\begin{equation}
\zeta=\tilde{\zeta}_{G}+\frac{3}{5}\tilde{f}_{NL}\left(\tilde{\zeta}_{G}^{2}-\tilde{\sigma}^{2}\right)=\tilde{h}(\tilde{\zeta}_{G}).
\label{localQuadInhomo}
\end{equation}
Therefore, both the amplitude and distribution of the small-scale perturbations are affected. In order to calculate the abundance of PBHs, the variables in equation (\ref{local fnl variables}) can then be inserted into equation (\ref{beta approx}).

\subsection{Cubic non-gaussianity, $g_{NL}$}
Here, we will follow the same steps as for $f_{NL}$, to show how the presence of a cubic term causes modal coupling. For this section, we will assume $f_{NL}=0$, and $\zeta$ to be given by
\begin{equation}
\zeta=\zeta_{G}+\frac{9}{25}g_{NL}\zeta_{G}^{3}.
\label{cubicNG}
\end{equation}
Again, using the peak-background split, one obtains:
\begin{equation}
\zeta=\left(1+\frac{27}{25}g_{NL}\zeta_{l}^{2}\right)\zeta_{s}+\left(\frac{27}{25}g_{NL}\zeta_{l}\right)\zeta_{s}^{2}+\left(\frac{9}{25}g_{NL}\right)\zeta_{s}^{3}+\mathcal{O}(\zeta_{l}),
\end{equation}
where again, the terms dependant only on $\zeta_{l}$ are neglected because they don't have a significant effect on PBH formation. The above expression can then be rewritten in terms of new variables $\tilde{\zeta}_{G}$, $\tilde{\sigma}$, $\tilde{f}_{NL}$ and $\tilde{g}_{NL}$, given by
\begin{equation}
\begin{split}
&\tilde{\zeta}_{G}=\left(1+\frac{27}{25}g_{NL}\zeta_{l}^{2}\right)\zeta_{s},\\
&\tilde{\sigma}=\left(1+\frac{27}{25}g_{NL}\zeta_{l}^{2}\right)\sigma_{s},\\
&\tilde{f}_{NL}=\left(\frac{9}{5}g_{NL}\zeta_{l}\right)\left(1+\frac{27}{25}g_{NL}\zeta_{l}^{2}\right)^{-2},\\
&\tilde{g}_{NL}=g_{NL}\left(1+\frac{27}{25}g_{NL}\zeta_{l}^{2}\right)^{-3}.\\
\end{split}
\label{bar gnl}
\end{equation}
Equation (\ref{cubicNG}) can then be rewritten as
\begin{equation}
\zeta=\tilde{\zeta}_{G}+\frac{3}{5}f_{NL}\left(\tilde{\zeta}_{G}^{2}-\tilde{\sigma}^{2}\right)+\frac{9}{25}\tilde{g}_{NL}\tilde{\zeta}_{G}^{3}.
\end{equation}
An expression for the abundance of PBHs in a given region of the universe, $\tilde{\beta}$, can then be derived as shown in section 2. 

In this section, it has been shown that long wavelength modes can affect the amplitude of local small scale perturbations and the non-gaussianity parameters, and in the next section the effect of this on the abundance of PBHs within a given region will be discussed.

\section{The isocurvature modes of PBH DM on CMB scales}

\begin{figure}[t]
\centering
\includegraphics[width=0.8\linewidth]{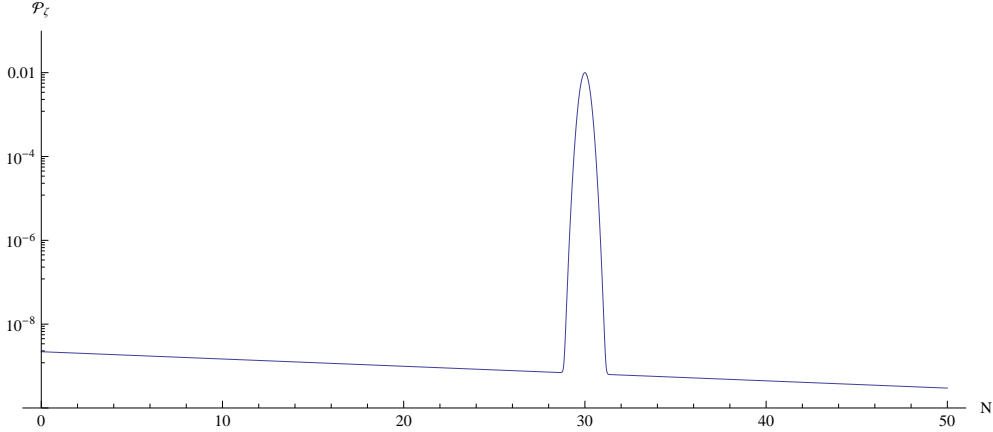}
\caption{An example of a power spectrum containing a narrow peak. $N$ represents number of e-folds, with smaller scales represented by larger $N$. The power spectrum is small on most scales with a spectral index of $n_{S}=0.96$, compatible with observations of the cosmic microwave background (CMB) and large scale structure (LSS). The narrow peak in the power spectrum corresponds to the scale at which will PBHs form.}
\label{narrow peak}
\end{figure}

The abundance of PBHs in a region of the universe can be affected significantly by large-scale curvature perturbation modes in different regions of the universe. If PBHs make up DM, then these differences in the abundance of PBHs will appear as fluctuations in the density of DM. In the presence of loacl-type non-gaussianity, the fluctuations in the DM can be significantly greater than the curvature perturbations responsible for producing them - and tight constraints can therefore be placed on the non-gaussianity parameters if this is the case from the isocurvature constraints from \emph{Planck}. 

We will define the difference in the abundance of PBHs at the time of formation, $\delta_{\beta}$, as
\begin{equation}
\delta_{\beta}=\frac{\beta-\bar{\beta}}{\bar{\beta}},
\label{deltabeta}
\end{equation}
where $\beta$ and $\bar{\beta}$ are the perturbed and background values of the PBH abundance at the time of formation respectively. If the large-scale curvature perturbation $\zeta$ is small, it can be related to $\delta_{\beta}$ by a constant factor $b$ (referred to the scale dependant bias in \cite{Tada:2015noa}),
\begin{equation}
\delta_{\beta}=b \zeta_{l},
\end{equation}
where $b$ is a function of the non-gaussianity parameters, the variance of the small-scale perturbations and the critical value for PBH formation $\zeta_{c}$. The factor $b$ therefore parameterises the bias of PBHs to form in the presence of large-scale curvature perturbations.

In this section, we will consider the case where the power spectrum is very small on all scales, except for a narrow region where there is a sharp spike -  which is responsible for the production of PBHs of a mass corresponding to this scale\footnote{The mass of a PBH is roughly equal to the horizon mass at the time of formation. See \cite{Young:2014ana} for further discussion.}. An example of such a power spectrum is given in figure \ref{narrow peak}. We therefore ignore in this section the presence of perturbations of intermediate scales, but extend the calculation in the following section to account for when there is a broad peak in the power spectrum.

The abundance of PBHs at a later time on a comoving slicing will be affected by difference in their density at the time of formation, as well as by the difference in expansion since the time of formation - in denser regions of the universe, inflation ends and PBHs form slightly later, so even if the PBH density is constant at the time of formation, the density will not be constant. To first order in $\zeta$, the density of PBHs can be expressed as
\begin{equation}
\Omega_{PBH}=\left(1+b\zeta+3\zeta\right)\bar{\Omega}_{PBH},
\end{equation}
where the $3\zeta$ term is simply the adiabatic mode expected from the expansion of the universe, and $\bar{\Omega}_{PBH}$ is the background density of PBHs. The $b\zeta$ term therefore represents a deviation from the expected amplitude of the mode if it was purely adiabatic - it is an isocurvature mode, which will either either be fully correlated, or fully anti-correlated depending on the sign of $f_{NL}$. If PBHs make up a significant fraction of the DM content of the universe, the constraints on isocurvature modes from \emph{Planck} can then be used to constrain $b$ - and therefore constrain the non-gaussianity parameters\footnote{Note that the reverse is also true - for a given value of the non-gaussianity parameters, an upper limit can be placed on the amount of DM which is made of PBHs}. For simplicity in this paper, except section 5.2, we will assume that DM is entirely composed of PBHs, and calculate corresponding constraints on the non-gaussianity parameters $f_{NL}$ and $g_{NL}$.
On CMB scales, the constraints from \emph{Planck} on isocurvature modes can be used \cite{Ade:2015lrj}
\begin{equation}
100\beta_{iso}=\begin{cases} 0.13 &\mbox{, fully correlated} \\
0.08 &\mbox{, fully anti-correlated},
\end{cases}
\end{equation}
where
\begin{equation}
\beta_{iso}=\frac{\mathcal{P}_{iso}}{\mathcal{P}_{iso}+\mathcal{P}_{\zeta}}.
\end{equation}
The fully correlated modes correspond to positive $b$, whilst fully anti-correlated corresponds to negative $b$ (and positive/negative $f_{NL}$ and $g_{NL}$ respectively). The isocurvature power spectrum is related to the curvature perturbation power spectrum as
\begin{equation}
\mathcal{P}_{iso}=b^{2}\mathcal{P}_{\zeta},
\end{equation}
and we therefore obtain constraints on $b$ as
\begin{equation}
-0.028<b<0.036.
\end{equation}
This result will now be used to derive a result on the non-gaussianity parameters.

\subsection{Isocurvature modes from $f_{NL}$}
In section 2, an expression for the abundance of PBHs at the time of formation $\beta$, was derived in terms of the non-gaussianity parameter $f_{NL}$, the variance of the gaussian component\footnote{$\sigma$ is related to the power spectrum as follows \cite{Byrnes:2007tm} 
\begin{equation} 
\mathcal{P}_{\zeta}=\sigma^{2}+\left(\frac{3}{5}\right)^2 \left( 4 f_{NL}^2+6 g_{NL}\right)\sigma^{4}\ln(kL)+ \left(\frac{3}{5}\right)^4 \left( 27 g_{NL}^2\right) \sigma^{6}\ln(kL)^{2}, 
\end{equation} 
where the higher order terms from $g_{NL}$ have also been included, and $\ln(kL)$ is a factor of around unity. Note that, since the non-gaussianity parameters are found to be very small, the higher order terms will not have a significant impact, and to a good approximation $\mathcal{P}_{\zeta}=\sigma^{2}$.} $\sigma^{2}$, and the critical value for collapse $\zeta_{c}$ - equation (\ref{beta approx}), with $h^{-1}$ given by equation (\ref{hfnl}). However, this calculation assumes there is no coupling to large scale modes (and is equivalent to the background value, $\bar{\beta}$, if large-scale perturbations are small - as is the case here). In section 3 it was shown how to account for the presence of a large scale modes - namely, by using the transformed variables $\tilde{f}_{NL}$ and $\tilde{\sigma}$ instead, given by equation (\ref{local fnl variables}) - which calculates the perturbed  abundance $\beta$.

By combing equations (\ref{hfnl}), (\ref{pos fnl beta}), (\ref{neg fnl beta}), (\ref{local fnl variables}) and (\ref{deltabeta}), it is possible to derive an expression for $\delta_{\beta}$ in terms of $f_{NL}$, $\sigma_{s}$ (where the $s$ subscript has been adopted to denote the small PBH scale), and the critical value $\zeta_{c}$. Expanding the expression to first order in $\zeta$ gives the result
\begin{equation}
\delta_{\beta}=\frac{25+30\zeta_{c}f_{NL}+36f_{NL}^{2}\sigma_{st}^{2}-5\sqrt{25+60\zeta_{c}f_{NL}+36f_{NL}^{2}\sigma_{s}^{2}}}{3f_{NL}\sigma_{s}^{2}\sqrt{25+60\zeta_{c}f_{NL}+36f_{NL}^{2}\sigma_{s}^{2}}}\zeta,
\end{equation}
and therefore $b$ is given by
\begin{equation}
b=\frac{25+30\zeta_{c}f_{NL}+36f_{NL}^{2}\sigma_{st}^{2}-5\sqrt{25+60\zeta_{c}f_{NL}+36f_{NL}^{2}\sigma_{s}^{2}}}{3f_{NL}\sigma_{s}^{2}\sqrt{25+60\zeta_{c}f_{NL}+36f_{NL}^{2}\sigma_{s}^{2}}},
\label{bfnl}
\end{equation}
or to first order in $f_{NL}$
\begin{equation}
b=\frac{6}{5}\left(1+\frac{\zeta_{c}^{2}}{\sigma_{s}^{2}}\right)f_{NL}.
\label{simple bfnl}
\end{equation}
As expected, a positive $f_{NL}$, which boosts the power spectrum on small scales in areas of higher density, produces a positive bias, and fully correlated isocurvature modes in PBH DM\footnote{The second expression for $b$ corresponds to equation (14) in \cite{Tada:2015noa}. The more complicated expression, equation (\ref{bfnl}), is because a gaussian distribution on small scales has not been assumed. The differences between the 2 calculations are discussed in Appendix B.}. Negative $f_{NL}$ has the opposite effect, and produces fully anti-correlated isocurvature modes.

\begin{figure}[t]
\centering
\begin{subfigure}{0.5\textwidth}
	\centering
	\includegraphics[width=\linewidth]{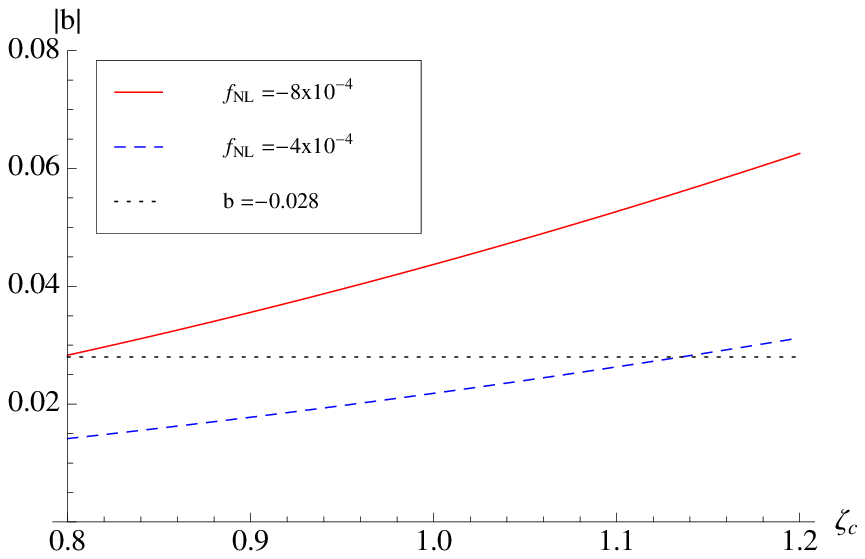}
\end{subfigure}%
\begin{subfigure}{0.5\textwidth}
	\centering
	\includegraphics[width=\linewidth]{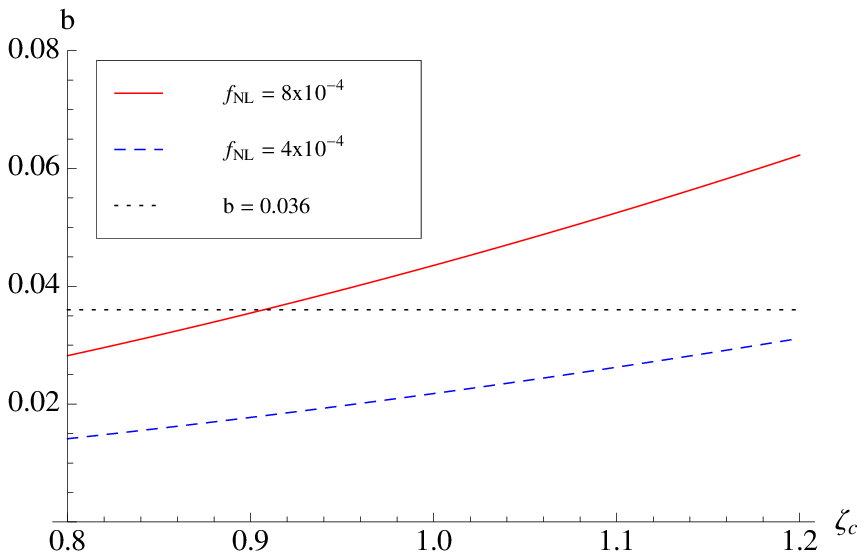}
\end{subfigure}
\caption{The plots above show the effects of a different threshold $\zeta_{c}$ on the PBH bias $b$ arising from an $f_{NL}$ term. A larger value of $\zeta_{c}$ suggests a larger bias factor. The left plot shows the effect for negative $f_{NL}$ and the right plot for positive $f_{NL}$. The dotted black lines represent the constrains on $b$ from the constraints on isocurvature modes from \emph{Planck}. $|f_{NL}|=8\times10^{-4}$ is typically excluded whilst $|f_{NL}|=4\times10^{-4}$ is typically allowed. To generate these plots the value $\sigma=0.15$ has been used.}
\label{zetac vs b fnl}
\end{figure}

\begin{figure}[t]
\centering
\begin{subfigure}{0.5\textwidth}
	\centering
	\includegraphics[width=\linewidth]{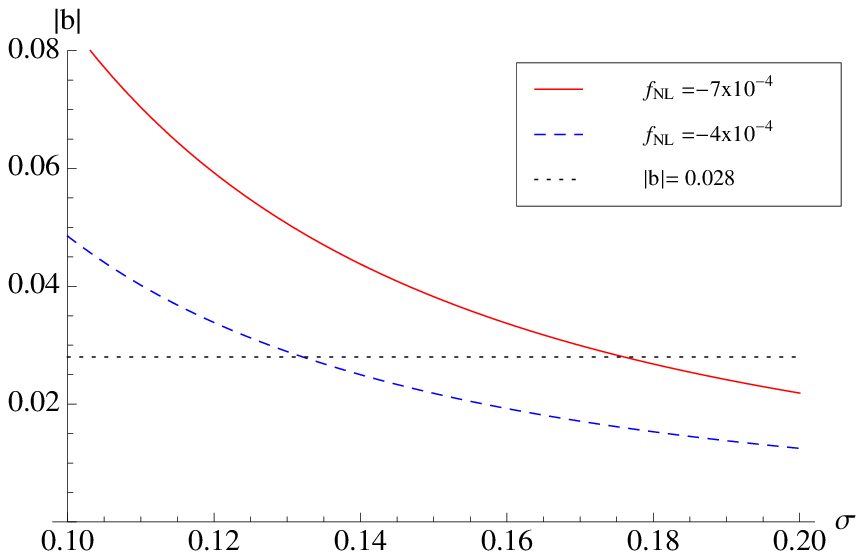}
\end{subfigure}%
\begin{subfigure}{0.5\textwidth}
	\centering
	\includegraphics[width=\linewidth]{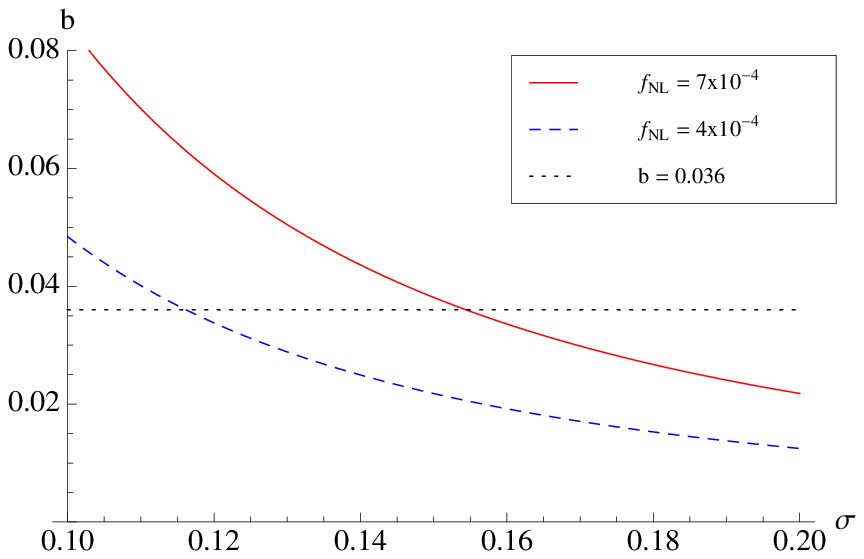}
\end{subfigure}
\caption{The effects of a different $\sigma$ on the PBH bias $b$ arising from an $f_{NL}$ term are investigated. A larger value of $\sigma$ suggests a smaller bias factor. The left plot shows the effect for negative $f_{NL}$ and the right plot for positive $f_{NL}$. The dotted black lines represent the constrains on $b$ from the constraints on isocurvature modes from \emph{Planck}. $|f_{NL}|=7\times10^{-4}$ is typically excluded whilst $|f_{NL}|=4\times10^{-4}$ is typically allowed. To generate these plots the value $\zeta_{c}=1$ has been used.}
\label{sigma vs b fnl}
\end{figure}

In order to investigate the constraints on the non-gaussianity parameters, it is necessary to estimate values for the other parameters involved, and how these would affect the constraints. The variance of the small scale perturbations and the critical value. 
\begin{itemize}
\item{First, $\zeta_{c}$ is considered: there is significant error in the exact value of the threshold value, due to uncertainty in the shape of the primordial perturbation which collapse to form PBHs. Most recent simulations have calculated the critical value in terms of the density contrast, finding $\delta_{c}\approx0.4$. This is consistent with the calculation here if the critical value of the curvature perturbation is related by a factor $\frac{4}{9}$, meaning $\zeta_{c}\approx 1$, which is consistent with the range of values found in figure \cite{Shibata:1999zs}. Figure \ref{zetac vs b fnl} shows how the factor $b$ depends on the critical value for different values of $f_{NL}$.}
\item{To calculate $\sigma_s$, it is necessary to first calculate the value of $\beta$ for which PBHs are otherwise unconstrained by observations and could be DM. The range of mass scales in which PBHs can form a significant fraction of DM is roughly $10^{17}\mathrm{g}<M_{PBH}<10^{24}\mathrm{g}$ \cite{Carr:2009jm}. The constraint on $\beta$ from the abundance of DM in this range are given by \cite{Josan:2009qn}
\begin{equation}
\beta<2\times10^{-19}\left(\frac{M_{PBH}}{f_{M}5\times10^{14}\mathrm{g}}\right)^{1/2},
\label{beta PBH mass}
\end{equation}
where $f_{M}$ is the fraction of the horizon mass which ends up inside the PBH\footnote{$f_{M}$ is a factor of order unity, which is neglected as it has very little effect on the calculated value of $\sigma_s$.}, and $M_{PBH}$ is the mass of the PBH. Assuming DM to be made up entirely of PBHs of a single mass scale within this range, $\beta$ can therefore range from $\beta<10^{-16}$ to $\beta<10^{-11}$. Assuming the most optimistic and pessimistic values for $\beta$ and $\zeta_{c}$, $\sigma_s$ is calculated to lie in the range $0.1<\sigma_s<0.2$ for close to gaussian perturbations \cite{Byrnes:2012yx}. Figure \ref{sigma vs b fnl} displays how $b$ changes with $\sigma_s$.}
\end{itemize}

Smaller values of the variance of the small scale perturbations, $\sigma_s^{2}$, would lead to tighter constraints on $f_{NL}$, whilst a smaller critical value $\zeta_{c}$ leads to tighter constraints on $f_{NL}$. Because a larger value of $\zeta_{c}$ implies a larger value of $\sigma_{s}$, these effects virtually cancel out - and the results presented below are therefore not sensitive to uncertainty in $\zeta_{c}$. 

Assuming PBH form at a single mass scale, the weakest constraint on $f_{NL}$ comes from considering the mass of the largest PBHs which could make up DM, which is taken to be $M_{PBH}=10^{25}$g, for which $\beta\approx10^{-14}$. If DM is made entirely of PBHs, the constraints on $f_{NL}$ are therefore
\begin{equation}
-4\times10^{-4}<f_{NL}<5\times10^{-4}.
\end{equation}
The results are not significantly different for PBHs of different mass. For example, for $M_{PBH}=10^{20}$g the constraints on $f_{NL}$ are
\begin{equation}
-3\times10^{-4}<f_{NL}<4\times10^{-4}.
\label{fnl-constraint}\end{equation}

\subsection{{Isocurvature modes from $g_{NL}$}}
In addition to $f_{NL}$, it is interesting to consider isocurvature modes arising from $g_{NL}$ and place constraints, or whether the effects of modal coupling from $g_{NL}$ could cancel the effects from $f_{NL}$. The effect of higher order terms are beyond the scope of this paper.

\begin{figure}[t]
\centering
	\includegraphics[width=0.6\linewidth]{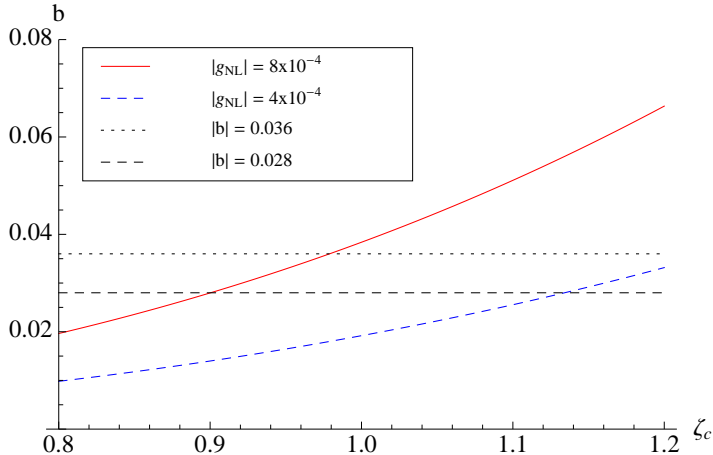}
\caption{The plots above show the effects of a different threshold $\zeta_{c}$ on the PBH bias $b$ arising from a $g_{NL}$ term. A larger value of $\zeta_{c}$ suggests a larger bias factor. As the expression for $b$, equation (\ref{bgnl}), is anti-symmetric under a change of sign of $g_{NL}$, the results for negative and positive $g_{NL}$ are shown on one plot - but with different constraints on the amplitude of $|b|$, represented by the dotted black lines. $|g_{NL}|=8\times10^{-4}$ is typically excluded whilst $|g_{NL}|=4\times10^{-4}$ is typically allowed. To generate these plots the value $\sigma=0.15$ has been used.}
\label{zetac vs b gnl}
\end{figure}

\begin{figure}[t]
\centering
	\includegraphics[width=0.6\linewidth]{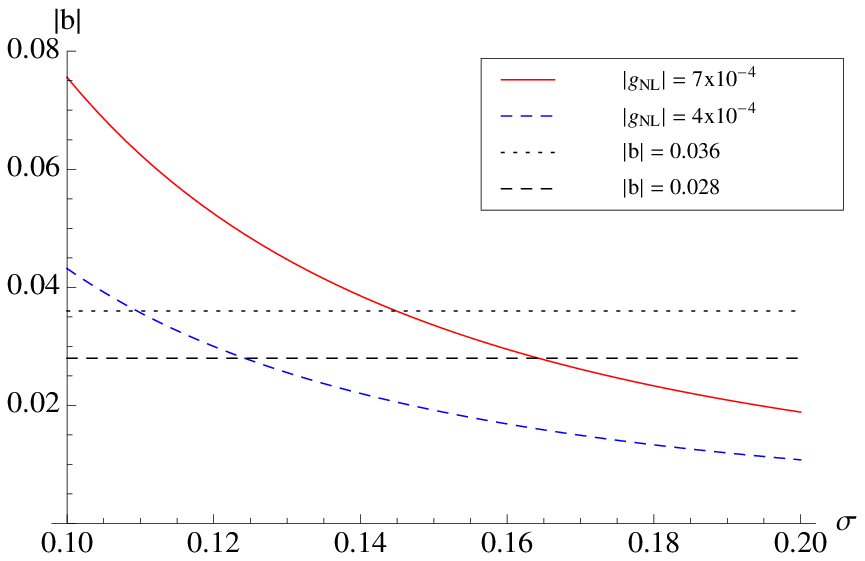}
\caption{This plot shows the effects of a different $\sigma$ on the PBH bias $b$ arising from a $g_{NL}$ term. A smaller value of $\sigma$ suggests a larger bias factor. As the expression for $b$, equation (\ref{bgnl}), is anti-symmetric under a change of sign of $g_{NL}$, the results for negative and positive $g_{NL}$ are shown on one plot - but with different constraints on the amplitude of $|b|$, represented by the dotted black lines. $|g_{NL}|=7\times10^{-4}$ is typically excluded whilst $|g_{NL}|=4\times10^{-4}$ is typically allowed. To generate these plots the value  $\zeta_{c}=1$ has been used. This range of $\sigma$ is used because it is approximately the range of values required to generate the correct number of PBHs to form DM (assuming that perturbations are close to gaussian).}
\label{sigmavsbgnl}
\end{figure}

The same derivation can be followed as that for $f_{NL}$, leading to an expression for $b$ to first order in $g_{NL}$
\begin{equation}
b=-\frac{27\left(\sigma_{s}^{2}-\zeta_{c}^{2}\right)\left(\sigma_{s}^{2}+\zeta_{c}^{2}\right)}{25\sigma_{s}^{2}\zeta_{c}}g_{NL}.
\label{bgnl}
\end{equation}
Again, as expected, positive $g_{NL}$ corresponds to fully correlated isocurvature modes, and negative $g_{NL}$ corresponds to fully anti-correlated isocurvature modes. The PBH bias factor $b$ is again a function of the non-gaussianity parameter $g_{NL}$, the variance of the small scale perturbations $\sigma_{s}^{2}$, and the formation threshold $\zeta_{c}$. The dependance of $b$ on $\zeta_{c}$ and $\sigma_{s}$ is shown in figures \ref{zetac vs b gnl} and \ref{sigmavsbgnl} respectively.

We see again that smaller values of $\sigma_s$ would lead to tighter constraints on $g_{NL}$, whilst a smaller $\zeta_{c}$ leads to tighter constraints on $g_{NL}$. However, unlike the case with $f_{NL}$, the constraint which can be placed on $g_{NL}$ depends on the value of $\zeta_{c}$, although only by a factor of $\mathcal{O}(10\%)$. The results presented below are the weakest constraints, corresponding to a low formation threshold, for PBHs of mass $10^{25}$g
\begin{equation}
-6\times10^{-4}<g_{NL}<7\times10^{-4}.
\end{equation}
Notice that these constraints are very comparable to those on $f_{NL}$, see (\ref{fnl-constraint}). The $f_{NL}$ term has an effect of $\mathcal{O}(10^{-5})$ on the small-scale power spectrum, whilst the $g_{NL}$ term only has an effect of $\mathcal{O}(10^{-10})$, and therefore, naively, the constraints on $g_{NL}$ would be expected to be roughly 5 orders of magnitude weaker than $f_{NL}$. However, a $g_{NL}$ term also has an effect on the small scale $\tilde{f}_{NL}$, as seen in equation (\ref{bar gnl}), of $\mathcal{O}(10^{-5})$, and because the abundance of PBHs is extremely sensitive to non-gaussianity, this causes significant isocurvature modes in the PBH DM. In the case where $\zeta_{l}=10^{-5}$ and $g_{NL}=10^{-3}$, then $\tilde{f}_{NL}\approx10^{-8}$. Such a small $\tilde{f}_{NL}$ nonetheless creates a perturbation in the PBH density of $\mathcal{O}(10^{-6})$, which represents an isocurvature mode of around $10\%$ of $\zeta$ - which is excluded by \emph{Planck}. Because the abundance of PBHs $\beta$ is sensitive to higher order non-gaussianity parameters \cite{Young:2013oia}, isocurvature modes are expected to rule out significant non-gaussianity at higher orders as well - although a quantitative calculation is beyond the scope of this paper. Higher order non-gaussianity parameters are considered briefly in section 5.4.

\section{Further consideration of constraints from isocurvature modes}
In section 4, constraints were placed separately on $f_{NL}$ and $g_{NL}$ separately, assuming that DM was entirely composed of primordial black holes. In this section, the calculation is extended to account for more general models.

\subsection{Isocurvature modes from $f_{NL}$ and $g_{NL}$}
The presence of non-zero non-gaussianity parameters has been shown to create significant isocurvature modes, which has led to very tight constraints on these parameters under the assumption that DM is composed entirely of PBHs. The calculation is now extended to account for non-zero $f_{NL}$ and $g_{NL}$ simultaneously - for example, it is possible that the effect of a large positive $f_{NL}$ and large negative $g_{NL}$ can cancel out, leaving a very small isocurvature mode.

Because the non-gaussianity parameters may now become quite large, the full numeric calculation for the PBH abundance is used to derive a value for the PBH bias $b$, for example by using equations (\ref{pos fnl beta}) or (\ref{neg fnl beta}) rather than the much simpler equation (\ref{beta approx}).

\begin{figure}[t]
\centering
\begin{subfigure}{0.49\textwidth}
	\centering
	\includegraphics[width=\linewidth]{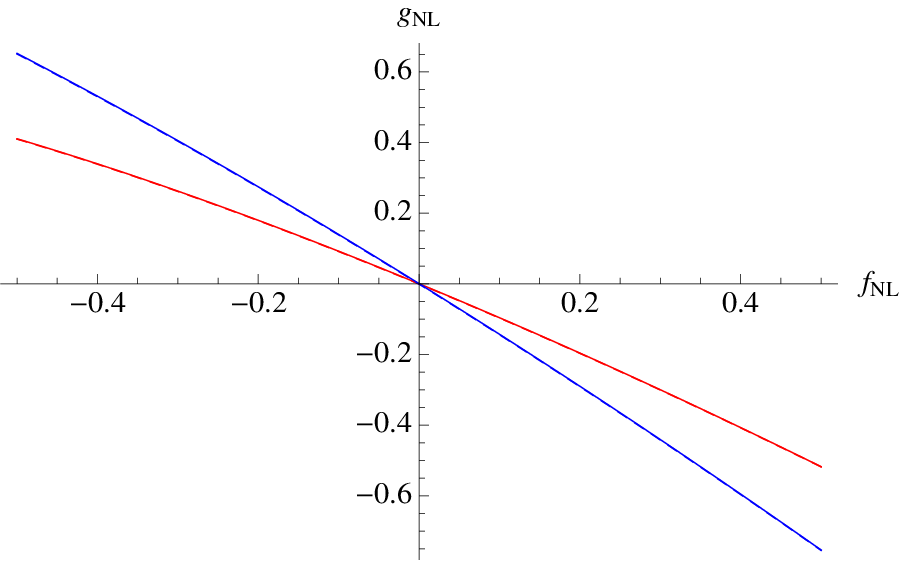}
\end{subfigure}%
\begin{subfigure}{0.49\textwidth}
	\centering
	\includegraphics[width=\linewidth]{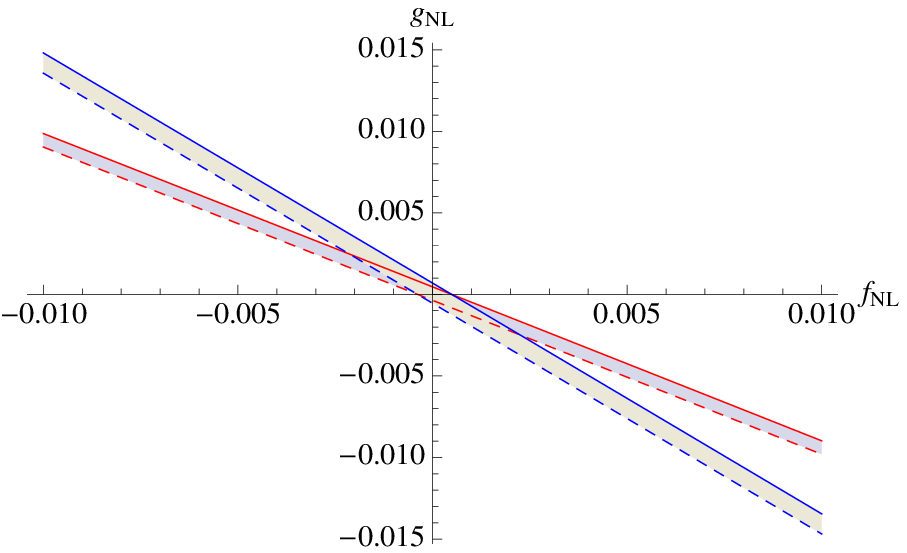}
\end{subfigure}
\caption{The constraints on simultaneous $f_{NL}$ and $g_{NL}$ are displayed. The right plot simply displays the central region of the plot on the left. The solid lines represent an upper limit from fully correlated isocurvature modes, whilst the dotted lines represent a lower limit from fully anti-correlated isocurvature modes. There is some uncertainty in the value of $g_{NL}$ given a value of $f_{NL}$ due to uncertainty in the critical value $\zeta_{c}$ - the blue lines are obtained using $\zeta_{c}=0.8$, and the red lines are obtained using $\zeta_{c}=1.2$. It can nonetheless be seen that large $f_{NL}$ or $g_{NL}$ are excluded unless very finely tuned. The shaded regions between the lines can be considered as $2\sigma$ contour plots from the \emph{Planck} constraints.}
\label{fnl vs gnl}
\end{figure}

Figure \ref{fnl vs gnl} shows the values of $g_{NL}$ that are permitted for different values of $f_{NL}$ for PBHs of mass $M_{PBH}=10^{25}$g. Whilst large values of $f_{NL}$ and $g_{NL}$ are allowed, there needs to be significant fine tuning to ensure that the resultant isocurvature modes are not excluded by the \emph{Planck} results - $g_{NL}$ needs to have the correct value to $\mathcal{O}(0.1\%)$. We note that there is some uncertainty in the value of $g_{NL}$ required for a given $f_{NL}$ due to the uncertainty in the formation threshold $\zeta_{c}$ - although this does not affect the conclusion that large non-gaussianity parameters are not allowed unless very very finely tuned. This conclusion is expected to remain true for higher-order terms \cite{Young:2013oia}.

\subsection{Fractional primordial black hole DM}
So far, it has been assumed that DM is made entirely of PBHs. The calculation is now extended to account for the fact that PBHs may only make up a small fraction of DM, and this is parameterised by $r_{PBH}$, the ratio of PBH density to DM density.
\begin{equation}
r_{PBH}=\frac{\Omega_{PBH}}{\Omega_{DM}}.
\end{equation}
In this case, the density of DM is described by
\begin{equation}
\Omega_{DM}=\left(1+r_{PBH}b\zeta+3\zeta\right)\bar{\Omega}_{DM},
\end{equation}
and the relative amplitude of the isocurvature modes is now given by $r_{PBH}b$. Therefore, from the \emph{Planck} constraints on isocurvature modes instead give constraints on the factor $r_{PBH}b$,
\begin{equation}
-0.028<r_{PBH}b<0.036.
\end{equation}

\begin{figure}[t]
\centering
\begin{subfigure}{0.5\textwidth}
	\centering
	\includegraphics[width=\linewidth]{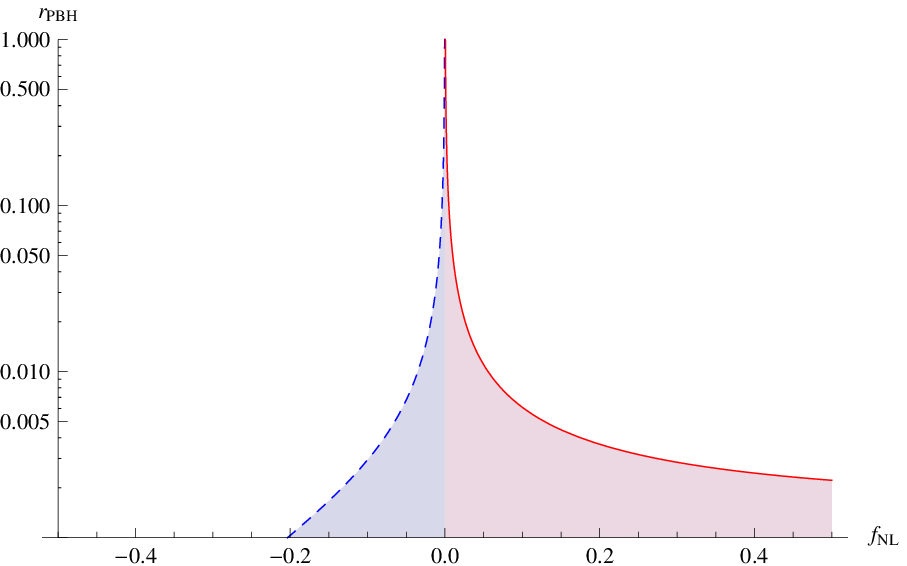}
\end{subfigure}%
\begin{subfigure}{0.5\textwidth}
	\centering
	\includegraphics[width=\linewidth]{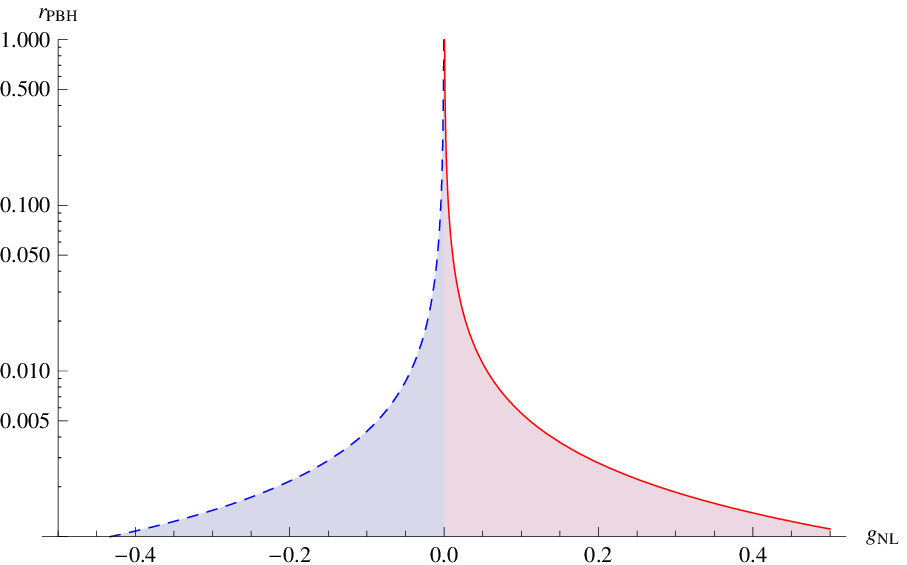}
\end{subfigure}
\caption{In the case where PBHs only make up a small fraction of the DM content of the universe, parameterised by $r_{PBH}$, the constraints on $f_{NL}$ and $g_{NL}$ can become significantly weaker. This is due to the fact a large isocurvature mode in the PBH density would only translate into a small isocurvature mode in the DM density. The plots above show the allowed values of $f_{NL}$ and $g_{NL}$ for different values of $r_{PBH}$. Whilst the plots show the constraints for PBHs of mass $M_{PBH}=10^{25}$g, the constraints are not very sensitive to the PBH mass.}
\label{fnlgnl vs rpbh}
\end{figure}

The constraints which can be placed on the non-gaussianity parameters therefore depend upon the PBH DM fraction, $r_{PBH}$. Figure \ref{fnlgnl vs rpbh} shows the allowed values of $f_{NL}$, $g_{NL}$ and $r_{PBH}$ if the PBH mass is $M_{PBH}=10^{25}$g. 
\begin{itemize}
\item{Large $r_{PBH}$: if PBHs make up a large fraction of DM then very tight constraints can be placed on the non-gaussianity parameters, $f_{NL},~g_{NL}<\mathcal{O}(10^{-2})$.}
\item{Small $r_{PBH}$: if PBHs make up a small fraction of DM, $r_{PBH}<0.1$, then the constraints on $f_{NL}$ and $g_{NL}$ weaken significantly. However, the non-gaussianity parameters only become larger than $1$ if $r_{PBH}<\mathcal{O}(10^{-3})$. In the case where $r_{PBH}$ is very small, the non-gaussianity parameters can become large and it is crucial to account for the effect of a non-gaussian distribution on the PBH forming scale, as done in this paper - as seen by the strong asymmetry for positive and negative $f_{NL}$.}
\end{itemize}

As $r_{PBH}$ becomes very small, $f_{NL}$ can become large and positive, but is still strongly restricted to not be large and negative. This is partly due to the fine tuning of the small scale power spectrum necessary to produce a small but not too large number of PBHs when $f_{NL}$ is negative - even a very small amount of modal coupling can mean that this fine tuning is disrupted in different regions of the universe, causing large amounts of variation in the number density of PBHs forming. This effect is not seen unless the non-gaussian distribution on small scales is accounted for. For $g_{NL}$, the constraints do not depend much on the sign of $g_{NL}$, and the small difference is due almost entirely to the difference in constraints from \emph{Planck} on fully, or fully anti-, correlated modes.

\subsection{Intermediate modes}

\begin{figure}[t]
\centering
\includegraphics[width=0.8\linewidth]{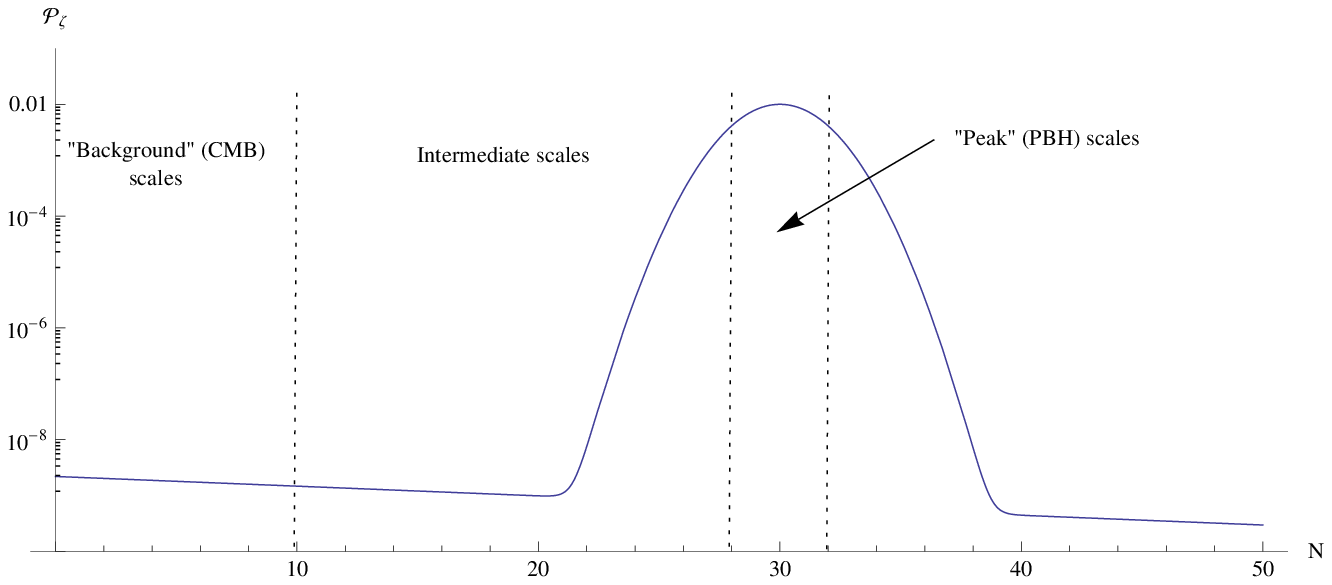}
\caption{An example of a power spectrum containing a broad peak. In this paper, there are 3 difference scales: the large "background" scales visible in the CMB, the small "peak" PBH forming scale (the exact scale of which depends on the mass PBH being considered), and the intermediate scales between the background and the peak. In such a case, the intermediate modes can have an effect on the PBH bias.}
\label{broad peak}
\end{figure}

The intermediate scales in between the large scales visible in the CMB and the small scale at which PBHs form have so far been ignored. This is a valid approximation if the power spectrum is small at all scales except for a narrow peak at the PBH forming scale, as in figure \ref{narrow peak}. However, this may not be the case if, for example, the power spectrum has a broad peak, as seen in figure \ref{broad peak}, or becomes blue at small scales. In this case, the abundance of PBHs, as well as the amplitude of isocurvature modes, can be significantly affected by the presence of perturbations on these intermediate modes.

If the power spectrum of the intermediate modes is not small, they will have a significant effect on the number of PBHs that form, as well as the isocurvature modes visible in PBH DM. This will be investigated in a similar to the peak-background split, and the curvature perturbation is split into short, intermediate, and long components:
\begin{equation}
\zeta_{G}=\zeta_{s}+\zeta_{i}+\zeta_{l}.
\end{equation}
The mass fraction of a given region of the universe going into PBHs is then calculated as before, as a function of $\zeta_{i}$ and $\zeta_{l}$, in addition to $f_{NL}$, $g_{NL}$, $\sigma_{s}$ and $\zeta_{c}$,
\begin{equation}
\beta=\beta\left(\zeta_{i},\zeta_{l}\right).
\end{equation}
However, the intermediate modes are too small scale to be observed in the CMB, and should therefore be averaged over:
\begin{equation}
\beta\left(\zeta_{l}\right)=\int\limits_{-\infty}^{\infty}\tilde{\beta}\left(\zeta_{i},\zeta_{l}\right)P(\zeta_{i})d\zeta_{i},
\end{equation}
where $\tilde{\beta}$ is the value of $\beta$ in different (intermediate-scale) regions of the universe, and $P(\zeta_{i})$ is the probability density function of $\zeta_{i}$, and is given by:
\begin{equation}
P(\zeta_{i})=\frac{1}{\sqrt{2Pi\langle\zeta_{i}^{2}\rangle}}\exp\left({-\frac{\zeta_{i}^{2}}{2\langle\zeta_{i}^{2}\rangle}}\right).
\end{equation}

In principle, $\langle\zeta_{i}^{2}\rangle$, can be obtained by integrating the power spectrum over the relevant range of scales. However, since this is unknown and model dependant, it is parameterised here by $r_{int}$, the ratio of the variance of intermediate modes $\langle\zeta_{i}^{2}\rangle$ to the variance of the short modes $\sigma_{s}^{2}$
\begin{equation}
r_{int}=\frac{\langle\zeta_{i}^{2}\rangle}{\sigma_{s}^{2}}.
\end{equation}

The value of $\langle\zeta_{i}^{2}\rangle$ can become larger than $\sigma_{s}^{2}$ due to the fact that many scales can contribute to $\zeta_{i}$, but only one scale contributes to $\zeta_{s}$. $\langle\zeta_{i}^{2}\rangle$ is calculated by integrating the power spectrum over the range of scales considered to be intermediate
\begin{equation}
\langle\zeta_{i}^{2}\rangle=\int\limits_{k_{min}}^{k_{max}}\frac{dk}{k}\mathcal{P}_{\zeta}(k),
\end{equation}
and can become large if the power spectrum is large over a significant range of this integration. In contrast, the PBH scale perturbations $\zeta_{s}$ are only composed of perturbations from one scale\footnote{Formally, $\sigma_{s}^{2}$ is given by integrating the power spectrum multiplied by a window function. However, provided that the spectral index is close to 1, or alternatively there is a peak spanning approximately 1 e-fold at the PBH scale, $\sigma_{s}^{2}$ is approximately equal to the power spectrum at that scale.}. Therefore, $\langle\zeta_{i}^{2}\rangle$ can become significantly larger than $\sigma_{s}^{2}$ even though the power spectrum has its largest value at the PBH scale. However, it is likely that in such a scenario, PBHs of multiple mass scales would be produced, which is discussed later.

\begin{figure}[t]
\centering
\begin{subfigure}{0.49\textwidth}
	\centering
	\includegraphics[width=\linewidth]{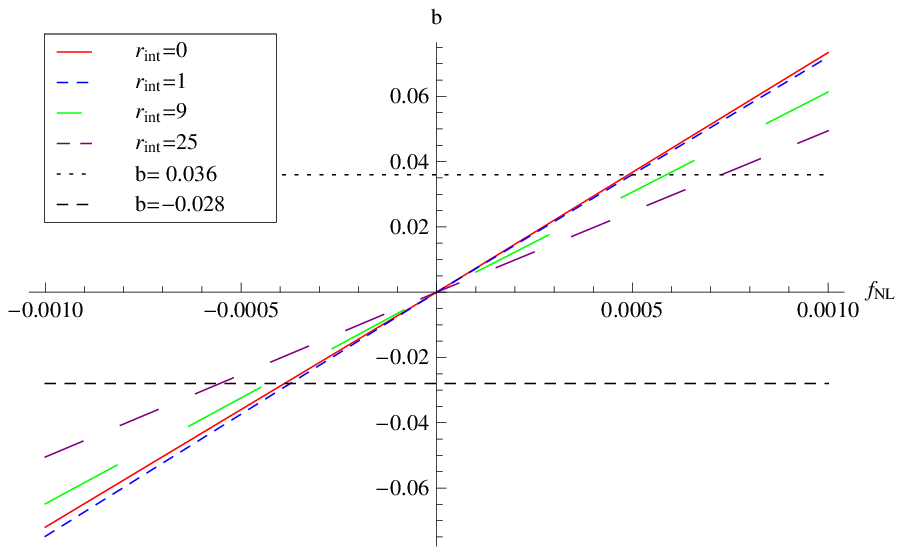}
\end{subfigure}%
\begin{subfigure}{0.49\textwidth}
	\centering
	\includegraphics[width=\linewidth]{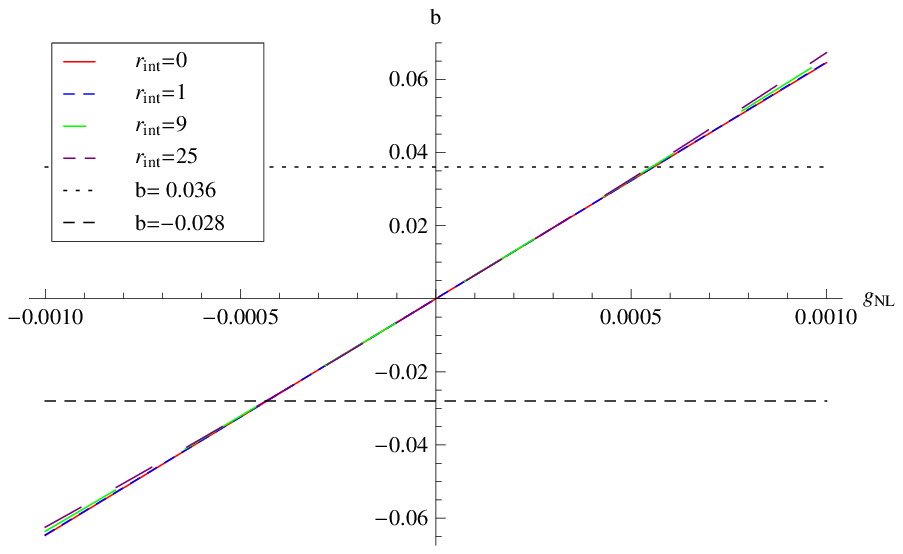}
\end{subfigure}
\caption{The effect of intermediate modes on the PBH bias $b$ is displayed for $M_{PBH}=10^{25}$g. The variance of the intermediate modes is parameterised by $r_{int}$, the ratio of $\langle\zeta_{int}^{2}\rangle$ to $\sigma_{s}^{2}$. The effect is negligible unless $r_{int}$ becomes large, in which case the PBH bias arising from an $f_{NL}$ term becomes significantly smaller, although has little effect for $g_{NL}$.}
\label{fnlgnl vs bint}
\end{figure}

The amplitude of the isocurvature modes therefore depends on the non-gaussianity parameters, the small scale power spectrum $\sigma_{s}^{2}$, the formation threshold $\zeta_{c}$, and $r_{int}$. A value for the PBH bias $b$ is then calculated numerically, figure \ref{fnlgnl vs bint} displays $b$ dependant on these variables. The effect of intermediate modes on the amplitude of isocurvature modes is relatively small for small $f_{NL}$ or $g_{NL}$ unless the variance of the intermediate scales is very large. The constraints on $f_{NL}$ can be weakened by a factor $\mathcal{O}(1)$, although the constraints on $g_{NL}$ are not significantly affected.

Note that a model where the power spectrum is large over a broad range of scales would likely also produce PBHs with a large range of masses, and vice versa. This fact does not affect the conclusions presented here, as the production of PBHs at all mass scales would be affected by bias in a similar way. We have shown that intermediate modes can significantly affect the PBH bias, although which modes are considered to be intermediate depends on the scale at which PBHs are forming, and therefore on the mass of PBHs forming. The exact constraints depend on the form of the power spectrum, and must therefore be calculated on a model by model basis, which goes beyond the scope of this paper - although the constraints will not be weaker than $f_{NL},~g_{NL}\lesssim10^{-3}$.

\subsection{Higher Order terms}
Whilst only the constraints on $f_{NL}$ and $g_{NL}$ have been calculated here, very tight constraints on higher order non-gaussianity parameters are also expected. In the same way that a $g_{NL}$ term has a small but significant effect on $f_{NL}$, equation (\ref{local NG}), higher order terms affect the previous term. Because the mass fraction of the universe forming PBHs is extremely sensitive to non-gaussianity parameters at higher orders \cite{Young:2013oia}, even very small changes to higher order non-gaussianity parameters due to modal coupling creates significant creates significant perturbations in the PBH density at large scales. As an example, we will consider a $5^{th}$ order term in local-type non-gaussianity:
\begin{equation}
\zeta=\zeta_{G}+\frac{81}{625}i_{NL}\zeta_{G}^{5}.
\end{equation}
Utilising the peak-background split gives a $4^{th}$ order term at small scales, $\tilde{h}_{NL}$, given by
\begin{equation}
\tilde{h}_{NL}=3i_{NL}\zeta_{l}.
\end{equation}
Inserting $\zeta_{l}\approx10^{-5}$ and $i_{NL}=10^{-3}$ gives $\tilde{h}_{NL}\approx10^{-8}$. The modulation of the $\tilde{h}_{NL}$ by the long wavelength mode $\zeta_l$ then generates a perturbation in the density of PBHs forming, $\delta_{\beta}\approx10^{-6}$. In the picture of PBH DM, this results in a fully-correlated isocurvature mode, with a bias factor of $b\approx0.1$ - which is excluded by Planck. Because it can be shown that high order terms have an effect on the preceding term which is linear in $\zeta$, tight constraints are expected on such non-gaussianity parameters, only weakening slightly as higher order terms are considered.

\section{Summary}
The effect of modal oupling under the presence of non-gaussianity of the local type produces significant isocurvature modes in the density of PBHs in the early universe. If PBHs make up a significant fraction of DM, the constraints on isocurvature modes in cold DM from \emph{Planck} can be used to constrain the non-gaussianity parameters - in this paper we have considered $f_{NL}$ and $g_{NL}$
Using the constraints from \emph{Planck} on isocurvature modes enables tight constraints to be placed on $f_{NL}$ and $g_{NL}$,
\begin{equation}
|f_{NL}|,|g_{NL}|<\mathcal{O}(10^{-3}),
\end{equation}
unless $f_{NL}$ and $g_{NL}$ have opposite signs and have been extremely finely tuned so that the effect from each term cancels. Cases where the constraints could become weaker have also been considered: if the power spectrum is large on scales between those visible in the CMB and the PBH forming scale, or if DM is only partially composed of PBHs, finding that under these conditions the constraints weaken very slightly (unless PBHs make up a very tiny fraction of DM). Therefore, the detection of significant numbers of PBHs would rule out significant local non-gaussianity, and vice versa. Our constraints are almost independent of the PBH mass, and can also be applied to Planck mass relics which may be left behind from the evaporation of small PBHs.


The production of isocurvature modes can therefore be used to constrain PBH forming models which may otherwise be permitted. For example, we will consider here two models which may be ruled out as mechanisms to produce PBH DM:
\begin{itemize}
\item{Hybrid inflation: hybrid inflation typically predicts a non-zero $f_{NL}$, but there is some freedom in the exact value. \cite{Clesse:2013jra} predicts $f_{NL}\approx -1/N_{*}$, where $N_{*}$ is the number of e-folds between horizon exit of some pivot scale and the end of horizon. Inflation is believed to have lasted at least $50-60$ e-folds, which would give $f_{NL}=\mathcal{O}(10^{-2})$ - several orders of magnitude higher than allowed by the constraints presented here. \cite{Mulryne:2011ni} predicts that $f_{NL}$ can span a range of values from $10^{-2}$ to $10^5$ - the entire range of which would be ruled out as a method of producing PBH DM.}
\item{The curvaton: the amount of non-gaussianity in the curvaton model depends on the density parameter, $\Omega_{\chi}$, of the curvaton, $\chi$, at the time it decays into radiation: $f_{NL}=-5/4$ if $\Omega_{\chi}=1$ \cite{Sasaki:2006kq}. Although higher order local non-gaussianity terms are generated, it is unlikely that these will generate small isocurvature perturbations to evade the constraints.}
\end{itemize}

There are, however, limitations to the calculations carried out in this paper. Notably, we have only considered local-type non-gaussianity, and throughout it has been assumed that $f_{NL}$ and $g_{NL}$ are scale invariant. We have also only calculated the dependance of isocurvature modes on $f_{NL}$ and $g_{NL}$, and shown them to a roughly equivalent effect - with $g_{NL}$ having only a marginally smaller effect. Higher order terms are therefore also likely to have a similar effect on isocurvature modes. We also note that it has recently been observed that sub-horizon perturbations at the time of PBH formation have an effect on whether a perturbation will collapse to form a PBH or not \cite{Nakama:2014fra}. The expected amplitude of these sub-horizon modes would be affected by modal coupling - and therefore affect the amount of PBHs forming, affecting the isocurvature modes. However, this effect is expected to be negligible whilst the non-gaussianity parameters are very small.

\section{Acknowledgements}
SY is supported by an STFC studentship, and CB is supported by a Royal Society University Research Fellowship. We thank David Seery, John Miller, Yuichiro Tada and Shuichiro Yokoyama for useful discussions.

\appendix
\section{Full expression for $\delta_{\beta}$ from a $g_{NL}$ term}
For completeness, the full expression for $\delta_{\beta}$  arising from a $g_{NL}$ term is included - though this expression is still only valid for small $g_{NL}$. This expression would replace the simpler equation (\ref{bgnl}).

\begin{equation}
\begin{split}
\delta_{\beta}=& \left(-50 3^{1/3} 10^{2/3}+\left(-27 \sqrt{g_{NL}} \text{$\zeta_{c}$}+\sqrt{300+729g_{NL} \text{$\zeta_{c}$}^2}\right)^{1/3} \left(45\ 3^{2/3} 10^{1/3} \sqrt{g_{NL}} \text{$\zeta_{c}$} \right.\right.\nonumber \\
& \left.\left.\left.-5\ 3^{1/6} 10^{1/3} \sqrt{100+243g_{NL} \text{$\zeta_{c}$}^2}+100 \left(-27 \sqrt{g_{NL}} \text{$\zeta_{c}$}+\sqrt{300+729g_{NL} \text{$\zeta_{c}$}^2}\right)^{1/3}\right.\right.\right.\nonumber \\
&  \left.\left.\left. -54g_{NL}\sigma^2 \left(-27 \sqrt{g_{NL}} \text{$\zeta_{c}$}+\sqrt{300+729g_{NL} \text{$\zeta_{c}$}^2}\right)^{1/3}\right)\right) \left(-25 3^{1/3} 10^{2/3} \sqrt{g^3 \left(100+243 g_{NL} \text{$\zeta_{c}$}^2\right)} \right. \right. \nonumber \\
& \left.\left. \left(-27 \sqrt{g_{NL}} \text{$\zeta_{c}$}+\sqrt{300+729 g_{NL} \text{$\zeta_{c}$}^2}\right)^{2/3}-250 \sqrt{3} g_{NL}^{3/2} \left(3^{1/6} 10^{1/3} \sqrt{100+243 g_{NL} \text{$\zeta_{c}$}^2}-20 \left(-27 \sqrt{g_{NL}} \text{$\zeta_{c}$} \right. \right. \right. \right. \nonumber \\
& \left. \left. \left. \left. +\sqrt{300+729 g_{NL} \text{$\zeta_{c}$}^2}\right)^{1/3}\right)+225 g_{NL}^2 \text{$\zeta_{c}$} \left(30\ 3^{1/6} 10^{1/3}-6 \sqrt{100+243 g_{NL} \text{$\zeta_{c}$}^2} \left(-27 \sqrt{g_{NL}} \text{$\zeta_{c}$} \right. \right. \right. \right. \nonumber \\
& \left. \left. \left. \left. +\sqrt{300+729 g_{NL} \text{$\zeta_{c}$}^2}\right)^{1/3}+3^{5/6} 10^{2/3} \left(-27 \sqrt{g_{NL}} \text{$\zeta_{c}$}+\sqrt{300+729 g_{NL} \text{$\zeta_{c}$}^2}\right)^{2/3}\right) \right. \right. \nonumber \\
& \left. \left. +243\ 3^{1/6} 10^{1/3} g_{NL}^3\sigma^2 \text{$\zeta_{c}$} \left(30+10^{1/3} \left(-81 \sqrt{g_{NL}} \text{$\zeta_{c}$}+3 \sqrt{300+729 g_{NL} \text{$\zeta_{c}$}^2}\right)^{2/3}\right) \right. \right. \nonumber \\
& \left. \left. -27\ 3^{1/3} g_{NL}^{5/2} \left(-450 3^{1/6} \text{$\zeta_{c}$}^2 \left(-27 \sqrt{g_{NL}} \text{$\zeta_{c}$}+\sqrt{300+729 g_{NL} \text{$\zeta_{c}$}^2}\right)^{1/3} \right. \right. \right. \nonumber \\
& \left. \left. +10^{1/3}\sigma^2 \sqrt{100+243 g_{NL} \text{$\zeta_{c}$}^2} \left(10\ 3^{1/3}+10^{1/3} \left(-27 \sqrt{g_{NL}} \text{$\zeta_{c}$}+\sqrt{300+729 g_{NL} \text{$\zeta_{c}$}^2}\right)^{2/3}\right)\right)\right) \nonumber \\
& \left(150\ 30^{1/3}\sigma^2 \sqrt{\frac{100}{g_{NL}}+243 \text{$\zeta_{c}$}^2} \left(-27 g_{NL}^2 \text{$\zeta_{c}$}+\sqrt{3} \sqrt{g^3 \left(100+243 g_{NL} \text{$\zeta_{c}$}^2\right)}\right)^{5/3} \left(-10 3^{1/3} \right. \right. \nonumber \\
& \left. \left. +10^{1/3} \left(-27 \sqrt{g_{NL}} \text{$\zeta_{c}$}+\sqrt{300+729 g_{NL} \text{$\zeta_{c}$}^2}\right)^{2/3}\right)\right)^{-1} \text{$\zeta$}.
\end{split}
\end{equation}

\section{Comparison with {\it ``Primordial black holes as biased tracers''}}
\begin{figure}[t]
\centering
\includegraphics[width=0.8\linewidth]{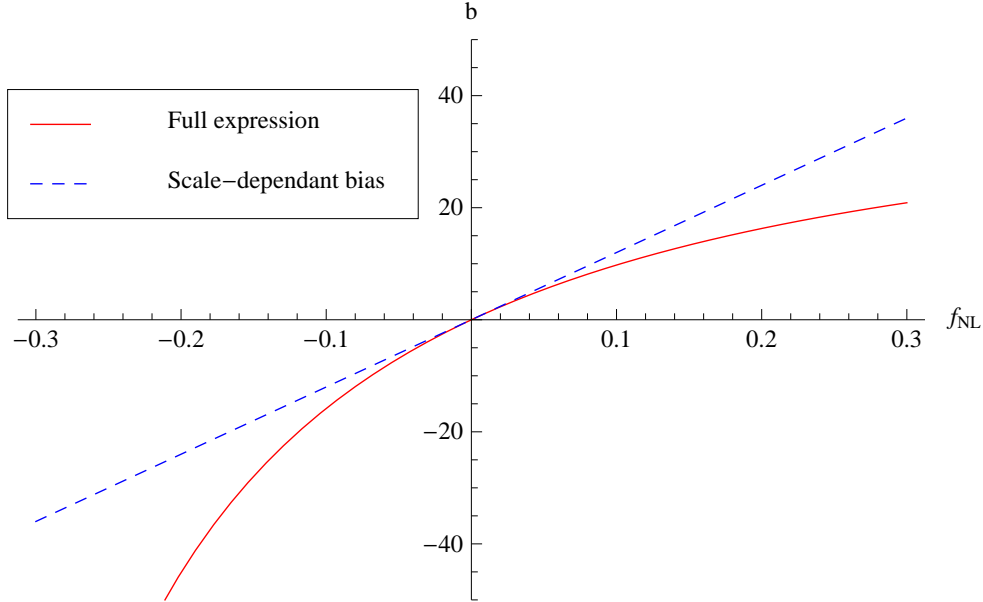}
\caption{A comparison of the results derived in this paper with those derived in \cite{Tada:2015noa}. The solid red line denotes the full expression for the PBH bias given by equation (\ref{bfnl}), and the dashed blue line represent the {\it scale-dependant bias} given by equation (14) in \cite{Tada:2015noa}. To make these plots, the values $\sigma_{s}=0.1$ and $\zeta_{c}=1$ have been used.}
\label{comparison}
\end{figure}

In their paper, {\it "Primordial black holes as biased tracers"} \cite{Tada:2015noa}, Tada and Yokoyama derive an expression for the {\it scale-dependant bias} given by
\begin{equation}
\Delta b(k)=2 f_{NL} \mathcal{M}_{l}^{-1}(k)\frac{\delta_{c}^{2}}{\sigma_{s}^{2}}.
\end{equation}
This is equivalent to equation (\ref{simple bfnl}) in this paper. The factor of $3/5$ difference is due to a different definition of $f_{NL}$, and the factor $\mathcal{M}_{l}^{-1}(k)$ is a result of their use of the density contrast rather than the curvature perturbation. The $+1$ in the brackets of equation (\ref{simple bfnl}) is a small correction and can be neglected. Therefore, the results for very small $f_{NL}$ in the 2 papers are equivalent. In figure \ref{comparison} the two expressions are compared. For $|f_{NL}|<\mathcal{O}(10^{-2})$ the two calculations match well, but diverge rapidly for larger $|f_{NL}|$.

It is therefore necessary to use the full calculation derived in this paper in situations where $f_{NL}$ could become larger than $10^{-2}$. Whilst such a large value of $f_{NL}$ is generally excluded by the constraints on isocurvature modes in the PBH DM scenario, it is relevant where higher order terms are considered, or that PBHs form a sub-dominant component of DM.

\bibliographystyle{JHEP}
\bibliography{bibfile}

\providecommand{\href}[2]{#2}\begingroup\raggedright\begin{thebibliography}{10}

\bibitem{Carr:2009jm}
B.~Carr, K.~Kohri, Y.~Sendouda, and J.~Yokoyama, {\it {New cosmological
  constraints on primordial black holes}},  {\em Phys.Rev.} {\bf D81} (2010)
  104019, [\href{http://xxx.lanl.gov/abs/0912.5297}{{\tt arXiv:0912.5297}}].

\bibitem{Pani:2014rca}
P.~Pani and A.~Loeb, {\it {Exclusion of the remaining mass window for
  primordial black holes as the dominant constituent of dark matter}},
  \href{http://xxx.lanl.gov/abs/1401.3025}{{\tt arXiv:1401.3025}}.

\bibitem{Capela:2014qea}
F.~Capela, M.~Pshirkov, and P.~Tinyakov, {\it {A comment on "Exclusion of the
  remaining mass window for primordial black holes ..."}},
  \href{http://xxx.lanl.gov/abs/1402.4671}{{\tt arXiv:1402.4671}}.

\bibitem{Defillon:2014wla}
G.~Defillon, E.~Granet, P.~Tinyakov, and M.~H. Tytgat, {\it {Tidal capture of
  primordial black holes by neutron stars}},  {\em Phys.Rev.} {\bf D90} (2014),
  no.~10 103522, [\href{http://xxx.lanl.gov/abs/1409.0469}{{\tt
  arXiv:1409.0469}}].

\bibitem{Carr:1994ar}
B.~J. Carr, J.~Gilbert, and J.~E. Lidsey, {\it {Black hole relics and
  inflation: Limits on blue perturbation spectra}},  {\em Phys.Rev.} {\bf D50}
  (1994) 4853--4867, [\href{http://xxx.lanl.gov/abs/astro-ph/9405027}{{\tt
  astro-ph/9405027}}].

\bibitem{Drees:2011hb}
M.~Drees and E.~Erfani, {\it {Running-Mass Inflation Model and Primordial Black
  Holes}},  {\em JCAP} {\bf 1104} (2011) 005,
  [\href{http://xxx.lanl.gov/abs/1102.2340}{{\tt arXiv:1102.2340}}].

\bibitem{Bugaev:2013fya}
E.~Bugaev and P.~Klimai, {\it {Axion inflation with gauge field production and
  primordial black holes}},  \href{http://xxx.lanl.gov/abs/1312.7435}{{\tt
  arXiv:1312.7435}}.

\bibitem{Bugaev:2011wy}
E.~Bugaev and P.~Klimai, {\it {Formation of primordial black holes from
  non-Gaussian perturbations produced in a waterfall transition}},  {\em
  Phys.Rev.} {\bf D85} (2012) 103504,
  [\href{http://xxx.lanl.gov/abs/1112.5601}{{\tt arXiv:1112.5601}}].

\bibitem{Lyth:2012yp}
D.~H. Lyth, {\it {The hybrid inflation waterfall and the primordial curvature
  perturbation}},  {\em JCAP} {\bf 1205} (2012) 022,
  [\href{http://xxx.lanl.gov/abs/1201.4312}{{\tt arXiv:1201.4312}}].

\bibitem{Halpern:2014mca}
I.~F. Halpern, M.~P. Hertzberg, M.~A. Joss, and E.~I. Sfakianakis, {\it {A
  Density Spike on Astrophysical Scales from an N-Field Waterfall Transition}},
   \href{http://xxx.lanl.gov/abs/1410.1878}{{\tt arXiv:1410.1878}}.

\bibitem{Lin:2012gs}
C.-M. Lin and K.-W. Ng, {\it {Primordial Black Holes from Passive Density
  Fluctuations}},  {\em Phys.Lett.} {\bf B718} (2013) 1181--1185,
  [\href{http://xxx.lanl.gov/abs/1206.1685}{{\tt arXiv:1206.1685}}].

\bibitem{Hotchkiss:2011gz}
S.~Hotchkiss, A.~Mazumdar, and S.~Nadathur, {\it {Observable gravitational
  waves from inflation with small field excursions}},  {\em JCAP} {\bf 1202}
  (2012) 008, [\href{http://xxx.lanl.gov/abs/1110.5389}{{\tt
  arXiv:1110.5389}}].

\bibitem{Linde:2012bt}
A.~Linde, S.~Mooij, and E.~Pajer, {\it {Gauge field production in supergravity
  inflation: Local non-Gaussianity and primordial black holes}},  {\em
  Phys.Rev.} {\bf D87} (2013), no.~10 103506,
  [\href{http://xxx.lanl.gov/abs/1212.1693}{{\tt arXiv:1212.1693}}].

\bibitem{Torres-Lomas:2014bua}
E.~Torres-Lomas, J.~C. Hidalgo, K.~A. Malik, and L.~A. Ure–a-L—pez, {\it
  {Formation of subhorizon black holes from preheating}},  {\em Phys.Rev.} {\bf
  D89} (2014) 083008, [\href{http://xxx.lanl.gov/abs/1401.6960}{{\tt
  arXiv:1401.6960}}].

\bibitem{Suyama:2014vga}
T.~Suyama, Y.-P. Wu, and J.~Yokoyama, {\it {Primordial black holes from
  temporally enhanced curvature perturbation}},  {\em Phys.Rev.} {\bf D90}
  (2014) 043514, [\href{http://xxx.lanl.gov/abs/1406.0249}{{\tt
  arXiv:1406.0249}}].

\bibitem{Green:2014faa}
A.~M. Green, {\it {Primordial Black Holes: sirens of the early Universe}},
  \href{http://xxx.lanl.gov/abs/1403.1198}{{\tt arXiv:1403.1198}}.

\bibitem{Jedamzik:1999am}
K.~Jedamzik and J.~C. Niemeyer, {\it {Primordial black hole formation during
  first order phase transitions}},  {\em Phys.Rev.} {\bf D59} (1999) 124014,
  [\href{http://xxx.lanl.gov/abs/astro-ph/9901293}{{\tt astro-ph/9901293}}].

\bibitem{Green:1997sz}
A.~M. Green and A.~R. Liddle, {\it {Constraints on the density perturbation
  spectrum from primordial black holes}},  {\em Phys.Rev.} {\bf D56} (1997)
  6166--6174, [\href{http://xxx.lanl.gov/abs/astro-ph/9704251}{{\tt
  astro-ph/9704251}}].

\bibitem{Josan:2009qn}
A.~S. Josan, A.~M. Green, and K.~A. Malik, {\it {Generalised constraints on the
  curvature perturbation from primordial black holes}},  {\em Phys.Rev.} {\bf
  D79} (2009) 103520, [\href{http://xxx.lanl.gov/abs/0903.3184}{{\tt
  arXiv:0903.3184}}].

\bibitem{Shandera:2012ke}
S.~Shandera, A.~L. Erickcek, P.~Scott, and J.~Y. Galarza, {\it {Number Counts
  and Non-Gaussianity}},  {\em Phys.Rev.} {\bf D88} (2013) 103506,
  [\href{http://xxx.lanl.gov/abs/1211.7361}{{\tt arXiv:1211.7361}}].

\bibitem{Ade:2015lrj}
{\bf Planck Collaboration} Collaboration, P.~Ade {\em et.~al.}, {\it {Planck
  2015. XX. Constraints on inflation}},
  \href{http://xxx.lanl.gov/abs/1502.0211}{{\tt arXiv:1502.0211}}.

\bibitem{Wands:2002bn}
D.~Wands, N.~Bartolo, S.~Matarrese, and A.~Riotto, {\it {An Observational test
  of two-field inflation}},  {\em Phys.Rev.} {\bf D66} (2002) 043520,
  [\href{http://xxx.lanl.gov/abs/astro-ph/0205253}{{\tt astro-ph/0205253}}].

\bibitem{Young:2014oea}
S.~Young and C.~T. Byrnes, {\it {The long-short wavelength mode coupling
  tightens primordial black hole constraints}},
  \href{http://xxx.lanl.gov/abs/1411.4620}{{\tt arXiv:1411.4620}}.

\bibitem{Maldacena:2002vr}
J.~M. Maldacena, {\it {Non-Gaussian features of primordial fluctuations in
  single field inflationary models}},  {\em JHEP} {\bf 0305} (2003) 013,
  [\href{http://xxx.lanl.gov/abs/astro-ph/0210603}{{\tt astro-ph/0210603}}].

\bibitem{Pajer:2013ana}
E.~Pajer, F.~Schmidt, and M.~Zaldarriaga, {\it {The Observed Squeezed Limit of
  Cosmological Three-Point Functions}},  {\em Phys.Rev.} {\bf D88} (2013),
  no.~8 083502, [\href{http://xxx.lanl.gov/abs/1305.0824}{{\tt
  arXiv:1305.0824}}].

\bibitem{Tanaka:2011aj}
T.~Tanaka and Y.~Urakawa, {\it {Dominance of gauge artifact in the consistency
  relation for the primordial bispectrum}},  {\em JCAP} {\bf 1105} (2011) 014,
  [\href{http://xxx.lanl.gov/abs/1103.1251}{{\tt arXiv:1103.1251}}].

\bibitem{Tada:2015noa}
Y.~Tada and S.~Yokoyama, {\it {Primordial black holes as biased tracers}},
  \href{http://xxx.lanl.gov/abs/1502.0112}{{\tt arXiv:1502.0112}}.

\bibitem{Niemeyer:1999ak}
J.~C. Niemeyer and K.~Jedamzik, {\it {Dynamics of primordial black hole
  formation}},  {\em Phys.Rev.} {\bf D59} (1999) 124013,
  [\href{http://xxx.lanl.gov/abs/astro-ph/9901292}{{\tt astro-ph/9901292}}].

\bibitem{Hawke:2002rf}
I.~Hawke and J.~Stewart, {\it {The dynamics of primordial black hole
  formation}},  {\em Class.Quant.Grav.} {\bf 19} (2002) 3687--3707.

\bibitem{Musco:2004ak}
I.~Musco, J.~C. Miller, and L.~Rezzolla, {\it {Computations of primordial black
  hole formation}},  {\em Class.Quant.Grav.} {\bf 22} (2005) 1405--1424,
  [\href{http://xxx.lanl.gov/abs/gr-qc/0412063}{{\tt gr-qc/0412063}}].

\bibitem{Musco:2008hv}
I.~Musco, J.~C. Miller, and A.~G. Polnarev, {\it {Primordial black hole
  formation in the radiative era: Investigation of the critical nature of the
  collapse}},  {\em Class.Quant.Grav.} {\bf 26} (2009) 235001,
  [\href{http://xxx.lanl.gov/abs/0811.1452}{{\tt arXiv:0811.1452}}].

\bibitem{Harada:2013epa}
T.~Harada, C.-M. Yoo, and K.~Kohri, {\it {Threshold of primordial black hole
  formation}},  {\em Phys.Rev.} {\bf D88} (2013), no.~8 084051,
  [\href{http://xxx.lanl.gov/abs/1309.4201}{{\tt arXiv:1309.4201}}].

\bibitem{Nakama:2013ica}
T.~Nakama, T.~Harada, A.~Polnarev, and J.~Yokoyama, {\it {Identifying the most
  crucial parameters of the initial curvature profile for primordial black hole
  formation}},  \href{http://xxx.lanl.gov/abs/1310.3007}{{\tt
  arXiv:1310.3007}}.

\bibitem{Shibata:1999zs}
M.~Shibata and M.~Sasaki, {\it {Black hole formation in the Friedmann universe:
  Formulation and computation in numerical relativity}},  {\em Phys.Rev.} {\bf
  D60} (1999) 084002, [\href{http://xxx.lanl.gov/abs/gr-qc/9905064}{{\tt
  gr-qc/9905064}}].

\bibitem{Young:2014ana}
S.~Young, C.~T. Byrnes, and M.~Sasaki, {\it {Calculating the mass fraction of
  primordial black holes}},  {\em JCAP} {\bf 1407} (2014) 045,
  [\href{http://xxx.lanl.gov/abs/1405.7023}{{\tt arXiv:1405.7023}}].

\bibitem{Byrnes:2012yx}
C.~T. Byrnes, E.~J. Copeland, and A.~M. Green, {\it {Primordial black holes as
  a tool for constraining non-Gaussianity}},  {\em Phys.Rev.} {\bf D86} (2012)
  043512, [\href{http://xxx.lanl.gov/abs/1206.4188}{{\tt arXiv:1206.4188}}].

\bibitem{Young:2013oia}
S.~Young and C.~T. Byrnes, {\it {Primordial black holes in non-Gaussian
  regimes}},  {\em JCAP} {\bf 1308} (2013) 052,
  [\href{http://xxx.lanl.gov/abs/1307.4995}{{\tt arXiv:1307.4995}}].

\bibitem{Byrnes:2007tm}
C.~T. Byrnes, K.~Koyama, M.~Sasaki, and D.~Wands, {\it {Diagrammatic approach
  to non-Gaussianity from inflation}},  {\em JCAP} {\bf 0711} (2007) 027,
  [\href{http://xxx.lanl.gov/abs/0705.4096}{{\tt arXiv:0705.4096}}].

\bibitem{Clesse:2013jra}
S.~Clesse, B.~Garbrecht, and Y.~Zhu, {\it {Non-Gaussianities and Curvature
  Perturbations from Hybrid Inflation}},  {\em Phys.Rev.} {\bf D89} (2014),
  no.~6 063519, [\href{http://xxx.lanl.gov/abs/1304.7042}{{\tt
  arXiv:1304.7042}}].

\bibitem{Mulryne:2011ni}
D.~Mulryne, S.~Orani, and A.~Rajantie, {\it {Non-Gaussianity from the hybrid
  potential}},  {\em Phys.Rev.} {\bf D84} (2011) 123527,
  [\href{http://xxx.lanl.gov/abs/1107.4739}{{\tt arXiv:1107.4739}}].

\bibitem{Sasaki:2006kq}
M.~Sasaki, J.~Valiviita, and D.~Wands, {\it {Non-Gaussianity of the primordial
  perturbation in the curvaton model}},  {\em Phys.Rev.} {\bf D74} (2006)
  103003, [\href{http://xxx.lanl.gov/abs/astro-ph/0607627}{{\tt
  astro-ph/0607627}}].

\bibitem{Nakama:2014fra}
T.~Nakama, {\it {The double formation of primordial black holes}},  {\em JCAP}
  {\bf 1410} (2014), no.~10 040, [\href{http://xxx.lanl.gov/abs/1408.0955}{{\tt
  arXiv:1408.0955}}].

\end{thebibliography}\endgroup

\end{document}